\documentclass[11pt,letterpaper]{article}

\usepackage[margin=1in]{geometry}
\usepackage[utf8]{inputenc}
\usepackage[T1]{fontenc}
\usepackage[expansion=false]{microtype}
\usepackage{amsmath, amssymb, amsthm}
\usepackage{mathtools}
\usepackage{graphicx}
\usepackage{booktabs}
\usepackage{tabularx}
\usepackage{multirow}
\usepackage{makecell}
\usepackage{enumitem}
\usepackage[hidelinks,colorlinks=true,linkcolor=blue!50!black,citecolor=blue!50!black,urlcolor=blue!50!black]{hyperref}
\usepackage{cleveref}
\usepackage{xcolor}
\usepackage{tikz}
\usetikzlibrary{positioning, arrows.meta, shapes.geometric, fit, backgrounds, calc, decorations.pathreplacing}
\usepackage{pgfplots}
\pgfplotsset{compat=1.18}
\usepackage{caption}
\usepackage{subcaption}
\usepackage{authblk}
\usepackage{natbib}
\usepackage{abstract}
\usepackage{titlesec}

\titlespacing*{\section}{0pt}{1.4em}{0.6em}
\titlespacing*{\subsection}{0pt}{1.0em}{0.4em}

\newtheorem{definition}{Definition}

\newtheorem{remark}{Remark}

\newcommand{\Prob}{\mathbb{P}}
\newcommand{\one}{\mathbb{1}}
\newcommand{\ILS}{\mathrm{ILS}}

\newcommand{\VPIN}{\mathrm{VPIN}}
\newcommand{\WN}{\mathrm{WN}}
\newcommand{\Topen}{T_{\mathrm{open}}}
\newcommand{\Tnews}{T_{\mathrm{news}}}
\newcommand{\Tres}{T_{\mathrm{resolve}}}
\newcommand{\Tevent}{T_{\mathrm{event}}}
\newcommand{\Tdead}{T_{\mathrm{deadline}}}

\title{\textbf{ForesightFlow:\\ An Information Leakage Score Framework\\ for Prediction Markets}}

\author[1]{Maksym Nechepurenko\thanks{Corresponding author: \texttt{maksym@devnull.ae}.}}
\affil[1]{Research Department, Devnull FZCO, Dubai, UAE}
\date{Preprint v2 (revised) --- \today}

\begin{document}

\maketitle

\begin{abstract}
\noindent
Decentralized prediction markets such as Polymarket aggregate dispersed beliefs into continuously updated price signals, but their on-chain transparency and pseudonymous participation also create an unusually fertile environment for informed trading on material non-public information. Recent empirical work has documented hundreds of millions of dollars in anomalous profits on Polymarket between 2024 and 2026; existing detection approaches are almost exclusively post-hoc and offer no actionable signal during the window when informed flow is moving prices.

We propose \textbf{ForesightFlow}, an information-theoretic framework for quantifying informed flow on prediction markets. We introduce the \emph{Information Leakage Score} (ILS), a label generator that quantifies how much of a market's terminal information move was priced in before the corresponding public news event, and show that ILS admits a clean interpretation in terms of the Murphy decomposition of the Brier score. We then specify the score's resolution-typology and operational scope conditions: ILS is interpretable only for event-resolved markets with substantive uncertainty at $T_{\text{open}}$, with article-derived rather than proxy-anchored news timestamps, and under an anchor-sensitivity robustness check. A pilot empirical study supports each scope condition: a resolution-anchored proxy for $T_{\text{news}}$ does not separate event-resolved markets from a matched control population; high ILS on high-consensus markets reflects a formula edge effect; and the proxy-based ILS distribution is not robust to anchor choice. A proof-of-concept article-derived $T_{\text{news}}$ recovery on the Epstein-files Barak market suggests that proxy quality is not the binding constraint and that wallet-level features rather than ILS alone are required to identify informed flow.

An audit of the publicly documented Polymarket insider-trading record reveals a structural finding: documented cases are systematically deadline-resolved (``Will event $X$ occur by date $Y$?''), falling outside the original ILS scope. We accordingly extend the score to a \emph{deadline-ILS} variant anchored to the underlying event timestamp, with a per-category exponential-hazard baseline for the time-to-event distribution. The extension closes the gap between the original methodology and the population in which informed trading has been empirically documented.

An end-to-end empirical evaluation of the deadline-ILS extension on the 2026 U.S.--Iran conflict cluster, including hazard-rate fits by category, a single-case ILS$^{\text{dl}}$ computation, cross-market wallet analysis, and the methodological refinements that the evaluation surfaces, is reported in a companion paper \citep{nechepurenko2026foresightflow_empirical}. We release the \emph{ForesightFlow Insider Cases} (FFIC) inventory, the full resolution-typology classification of the 911{,}237-market corpus, and all system code openly at \url{https://github.com/ForesightFlow}.
\end{abstract}



\vspace{1em}
\hrule
\vspace{1em}

\section{Introduction}
\label{sec:intro}

Prediction markets are mechanisms for aggregating dispersed private information into a publicly observable price signal \citep{wolfers2004prediction, arrow2008promise}. By allowing participants to trade contracts whose payoffs are contingent on real-world outcomes, these markets incentivize honest revelation of beliefs and convert heterogeneous private judgements into a single consensus probability estimate \citep{surowiecki2004wisdom}. Modern blockchain-based platforms such as Polymarket and Kalshi have brought these mechanisms to scale: Polymarket alone processed over five hundred million dollars of contract volume in 2024 \citep{polymarket2024volume}, and weekly volume across the sector reached the billion-dollar range during peaks of the 2024 U.S. presidential election cycle. The published probabilities of these markets have, in turn, become reference points that journalists, donors, traders, and institutions cite when deciding how to respond to political and economic uncertainty---a coordination role that is structurally distinct from forecasting accuracy and that creates its own demand for signal integrity \citep{nechepurenko2026focal, manski2006interpreting}.

The same features that make these markets useful information aggregators---low-friction global access, on-chain transparency, and pseudonymous participation---also create unusually permissive conditions for trading on material non-public information (MNPI). Unlike traditional securities and derivatives venues, where surveillance, registration, and statutory anti-fraud regimes constrain informed participants, decentralized prediction markets operate in a regulatory gray area in which classical insider-trading prohibitions map only awkwardly onto the underlying conduct \citep{mitts2026iran}. Documented case studies illustrate the resulting opportunity space: in the hours before the February 2026 U.S.--Israeli strike on Iran, six newly created Polymarket wallets purchased ``Yes'' shares at prices as low as ten cents and collectively realized profits of roughly \$1.2 million when the market resolved hours later. Comparable episodes surround the December 2025 release of Google's proprietary ``Year in Search'' rankings, the public announcement of Taylor Swift's engagement, the U.S. operation in Venezuela, and several recent OpenAI product launches. \citet{mitts2026iran} estimate that approximately \$143 million in aggregate anomalous profit was extracted across Polymarket during a two-year window, drawing on a composite statistical screen applied to over 210{,}000 wallet--market pairs. \citet{imdea2025polymarket} independently documented \$40 million of arbitrage and quasi-arbitrage profit from a complementary on-chain measurement.

These findings establish that informed trading is a quantitatively significant phenomenon in decentralized prediction markets. They do not, however, address what we view as the operationally relevant question: \emph{can informed flow be detected while a market is still open}, in time for an outside observer to either act on the signal or, in a regulatory or platform-integrity setting, to intervene before resolution? Existing work is overwhelmingly post-hoc. \citet{mitts2026iran} use a five-feature screen that explicitly conditions on profitability and proximity to resolution, both of which are observable only after the fact. \citet{imdea2025polymarket} estimate realized arbitrage profits using settled outcomes. Commercial trackers such as Polysights and Wallet Master expose ex-post wallet performance but do not translate it into forward-looking signals. The gap between post-hoc identification and real-time detection is substantial, both technically and operationally. Real-time detection requires that the relevant features---price action, order flow, wallet behavior, news context---can be computed from information available before resolution, and that the resulting score is calibrated against an outcome label that is itself recovered from historical data without leakage.

The stakes of this gap are amplified by what \citet{nechepurenko2026focal} terms the \emph{authority paradox}: prediction markets tend to acquire social credibility---being widely cited, organizing donor and elite behavior, shaping media narrative---precisely during the periods when their underlying microstructure is least mature, with thin liquidity and high price impact. Informed flow that arrives during these windows is most consequential not because it generates the largest profits, but because it shapes the public signal at the exact moment when that signal carries the most coordination weight. Real-time detection is therefore not merely a trading concern; it is integral to the broader question of when and how decentralized markets can be trusted as public information infrastructure.

\subsection{Contribution}
\label{sec:contribution}

This paper develops the methodological foundations of \textbf{ForesightFlow}, a system whose design goal is real-time detection of informed flow in active prediction markets, with primary application to Polymarket. Our contributions are as follows.

\begin{enumerate}[leftmargin=1.4em, itemsep=0.3em]
\item \textbf{Reformulation.} We separate the problem of \emph{retrospective forensics} from the problem of \emph{real-time detection}, and argue that the latter is the substantively interesting case for trading, regulation, and platform design. The reformulation has architectural consequences: features must be causal in the information they use, and labels must be generated from resolved markets in a manner that does not contaminate the training-time covariates of the detector.

\item \textbf{Information Leakage Score (ILS).} We introduce a scalar label generator that, for any resolved binary market, quantifies the fraction of the total information move that occurred before the first public mention of the resolution-relevant event. We give a formal definition (\Cref{sec:ils}) together with three explicit scope conditions---an edge-effect condition on the opening price, a trivial-resolution threshold on the total information move, and an anchor-sensitivity condition that requires the recovered value to be robust across multiple anchor specifications.

\item \textbf{Resolution typology.} We introduce a three-way classification of resolved Polymarket markets---event-resolved, deadline-resolved, and unclassifiable---that turns out to be necessary for a principled application of the score. The classification is empirically validated by the structural fact that markets the classifier identifies as deadline-resolved exhibit a 100\% NO-resolution rate by construction (\Cref{sec:ils-typology}).

\item \textbf{Murphy decomposition of leakage.} We connect ILS to the classical Murphy decomposition of the Brier score \citep{murphy1973decomposition}, showing that pre-news leakage corresponds to front-loaded resolution---the discriminative component of forecasting performance, accumulated before the public information set has caught up. This provides a principled bridge between the proper-scoring-rule literature and the on-chain forensics literature, and connects our work to the parallel evaluation framework developed by \citet{nechepurenko2026foresight}.

\item \textbf{Microstructure adaptation.} We adapt three families of classical market-microstructure measures of informed trading---PIN \citep{easley1996liquidity}, VPIN \citep{easley2012vpin}, and Kyle's lambda \citep{kyle1985continuous}---to the specific setting of discrete, bounded-price binary outcome markets traded on a hybrid on-chain CLOB (\Cref{sec:micro}). The adaptations are non-trivial: classical measures rely on continuous mid-quote dynamics and on the absence of explicit trade-direction information, both of which differ substantially from the Polymarket case. We further connect these measures to the Signal Credibility Index of \citet{nechepurenko2026focal}, which uses a closely related family of microstructure diagnostics for a different downstream purpose---measuring the social authority of a price move rather than its informational provenance.

\item \textbf{On-chain wallet features.} We define a wallet-novelty score that complements microstructure features by exploiting the public on-chain identity layer that decentralized platforms expose by design. The score is a weighted composite of wallet age, prior market participation, funding-source concentration, and entry timing relative to resolution.

\item \textbf{Category-conditional priors.} We restrict the empirical scope of the work to three categories where informed trading is plausible \emph{a priori} and where documented cases exist: military and geopolitical actions, corporate proprietary disclosures, and regulatory decisions. Sports, weather, and election-polling markets serve as null-hypothesis controls. The resulting category-conditional priors form an attention layer that focuses real-time monitoring on the markets where informed flow is most likely to appear.

\item \textbf{ForesightFlow Insider Cases (FFIC) inventory.} We release a curated validation set mapping eight publicly documented episodes of suspected informed trading on Polymarket---across military, regulatory, and corporate categories---to concrete market identifiers, event timestamps, and references. The inventory spans markets ranging from \$148{,}000 to \$2.57{,}billion in traded volume, and includes structural patterns that are themselves informative: in particular, the observation that documented insider activity does not concentrate at the highest-volume markets. The inventory is released under CC-BY-4.0 as a community resource for reproducibility (\Cref{sec:data-validation}).

\item \textbf{Pilot empirical study with negative findings.} We report a pilot ILS computation on $n = 725$ event-resolved markets together with a matched control population, a proxy-sensitivity analysis at four anchor offsets, a deep dive on the highest-signal Epstein-files cluster, and a single-market article-derived $\Tnews$ proof-of-concept. The findings tighten the methodology in several ways and identify a structural gap between the score's scope and the population in which informed trading has been documented (\Cref{sec:pilot}).

\item \textbf{Deadline-ILS extension.} Motivated by the pilot's structural finding, we formalize an extension of ILS to deadline-resolved markets that anchors the score to a baseline at $\Topen$, defines a parallel formulation for the YES-resolved and NO-resolved cases, adopts a per-category constant-hazard model for the time-to-event distribution, and inherits the existing scope conditions with one explicit modification (\Cref{sec:extension}). The end-to-end empirical evaluation of the extension is the subject of the companion paper \citep{nechepurenko2026foresightflow_empirical}.

\item \textbf{System architecture.} We describe the data pipeline, scoring engine, and detection layer of a production system that ingests Polymarket data, news from GDELT \citep{leetaru2013gdelt}, and on-chain wallet data, and produces a calibrated detection probability for each active market in the relevant categories. All system code and live deployment artifacts are released openly at \url{https://github.com/ForesightFlow} and \url{https://foresightflow.xyz}.
\end{enumerate}

\subsection{Scope and limitations of this draft}

This is a working preprint. The methodology and infrastructure (\Cref{sec:related,sec:data,sec:ils,sec:micro}) are complete; an empirical pilot study with explicit negative findings (\Cref{sec:pilot}) and a methodological extension that the pilot motivated (\Cref{sec:extension}) are likewise complete in their present form. The remaining empirical work---in particular, the deadline-ILS computation across the FFIC validation inventory, the trade-history backfill for the largest-volume insider-case markets, and the full detector training and backtest---is the subject of subsequent revisions of this work. We note throughout where empirical claims are conditional on infrastructure that is in progress or where the next-stage evaluation is still pending.

\subsection{Outline}

The remainder of this paper is organized as follows. \Cref{sec:related} reviews related literature on market microstructure, prediction-market efficiency, and on-chain forensics, and positions ForesightFlow within this body of work. \Cref{sec:data} describes the data sources, the operational definitions of the three target categories, the methodology for recovering news timestamps, the structure of the historical sample, and the FFIC validation inventory. \Cref{sec:ils} formalizes the Information Leakage Score, its auxiliary metrics, the scope conditions under which the score is interpretable, and the resolution-typology refinement. \Cref{sec:micro} details the microstructure and wallet-feature adaptations and outlines the detector architecture. \Cref{sec:pilot} reports a pilot empirical study whose negative findings tighten the methodology, a single-market article-derived $\Tnews$ proof-of-concept that further refines the operational reading of the framework, and a structural finding from an attempt to scale the proof-of-concept to the full FFIC inventory: the documented cases that motivate the validation set are systematically deadline-resolved, falling outside the score's current scope. \Cref{sec:extension} formalizes the extension of ILS to deadline-resolved cases. The end-to-end empirical evaluation of the extension is the subject of the companion paper \citep{nechepurenko2026foresightflow_empirical}; \Cref{sec:summary} summarizes system design and identifies the next-stage empirical work that flows from both papers jointly.

\section{Related Work}
\label{sec:related}

ForesightFlow sits at the intersection of three established literatures: market microstructure and informed-trading detection, prediction-market design and efficiency, and blockchain forensics on decentralized exchanges. We review each, and then state explicitly the gap that the present work addresses.

\subsection{Market microstructure and informed trading}
\label{sec:related-micro}

The modern theoretical treatment of informed trading begins with \citet{kyle1985continuous}, who modelled price formation as the equilibrium between a single informed trader, noise traders, and a competitive market maker, and derived a closed-form expression---now universally known as Kyle's lambda---for the per-unit price impact of order flow. \citet{glosten1985bid} provided a complementary microstructural account in which the bid--ask spread itself reflects the market maker's expected loss to informed counterparties.

\citet{easley1996liquidity, easley1997information, easley2002information} translated these theoretical insights into an empirical procedure, the \emph{Probability of Informed Trading} (PIN) measure. PIN models the daily arrival of orders as a mixture of informed and uninformed flows and infers the informed fraction by maximum likelihood from the joint distribution of buy and sell counts. PIN has been applied across a wide range of markets and has spawned a substantial methodological literature; \citet{duarte2009pin} document a number of practical estimation difficulties, and \citet{easley2012vpin} introduce VPIN, a volume-synchronized variant that replaces calendar time with volume buckets and is more robust to changes in trading intensity. VPIN attracted both wide adoption and pointed criticism \citep{andersen2014vpin}, and the resulting debate has clarified both the strengths and limitations of toxicity measures derived from order-flow imbalance.

For our purposes, the specific form of PIN, VPIN, or Kyle's lambda matters less than the underlying observation: in the presence of informed traders, the joint distribution of order flow and price change deviates systematically from what is observed under uninformed-only conditions, and these deviations are detectable in real time from order-book data alone. The classical literature has established this for continuous-quote equity and futures markets. The adaptation to discrete-outcome, hybrid on-chain CLOB markets, where prices are bounded in $[0, 1]$ and trade direction is explicitly recorded on-chain, is to our knowledge new.

\subsection{Prediction market design and efficiency}
\label{sec:related-pm}

The design and efficiency properties of prediction markets are surveyed in \citet{wolfers2004prediction} and \citet{arrow2008promise}. \citet{hanson2003combinatorial, hanson2007logarithmic} introduced the logarithmic market scoring rule (LMSR), a subsidy-based market-maker mechanism that guarantees bounded operator loss while maintaining incentives for honest probability revelation. The Iowa Electronic Markets demonstrated empirically that market-aggregated probabilities can match or exceed traditional polling for political forecasting. More recent work continues to find that prediction markets are well-calibrated on average across many domains \citep{atanasov2017distilling}, providing the baseline against which deviations associated with informed trading can be measured.

A separate strand of work studies the boundaries of market efficiency in this setting. \citet{grossman1980impossibility} pointed out that perfectly informative prices cannot coexist with costly information acquisition, implying that informed traders earn rents proportional to their information advantage and that some level of mispricing must persist in equilibrium. In the prediction-market context this is the operational space within which informed flow is profitable. \citet{schoenegger2024large} have shown that ensembles of large language models can approach human crowd accuracy on prediction-market questions, raising the related question of how machine forecasters interact with the same price signal.

The question of whether decentralized prediction markets are efficient in the strong sense---i.e., whether prices fully reflect all relevant information including private information held by insiders---is, by the case-study evidence summarized in \Cref{sec:intro}, empirically negative. The interesting question is therefore not whether informed trading occurs but how much of the eventual information move it accounts for and whether its signature is detectable before resolution.

\subsection{Blockchain forensics and on-chain measurement}
\label{sec:related-onchain}

The transparency of on-chain settlement enables a class of empirical analyses that has no exact counterpart in traditional financial markets. \citet{daian2020flash} introduced the concept of Miner Extractable Value (MEV), demonstrating that ordering, insertion, and censorship of transactions in decentralized exchanges constitute a form of latency arbitrage with measurable economic consequences. The ``Flash Boys 2.0'' framing connected this directly to the equity-market high-frequency trading literature \citep{budish2015hft}.

In the prediction-market setting specifically, two recent papers anchor the empirical baseline that the present work builds on. \citet{imdea2025polymarket} analyzed 86 million Polymarket bets between April 2024 and April 2025 and identified two structural arbitrage classes---intra-market rebalancing and inter-market combinatorial---accounting for roughly \$40 million of realized profit. \citet{mitts2026iran} extended the analysis to informed trading specifically, screening over 93{,}000 markets and 210{,}000 wallet--market pairs through a composite score combining cross-sectional bet size, within-trader bet size, profitability, pre-event timing, and directional concentration. They report a 69.9\% win rate on flagged trades, exceeding the null distribution by more than 60 standard deviations under a permutation test, and estimate aggregate anomalous profit of approximately \$143 million.

Two features of these prior studies are important for situating our contribution. First, both rely on resolved markets and on features (such as profitability) that are observable only after resolution. Second, both treat the question as one of \emph{statistical detection over a population of markets and wallets} rather than \emph{real-time inference on an individual active market}. A third feature, important for the methodological discussion in \Cref{sec:pilot-ffic-audit,sec:extension}, is that the canonical case studies highlighted by \citet{mitts2026iran}---the October 2024 Iran-strike market, the 2026 U.S.--Iran cluster, and several regulatory-timing markets---are structurally deadline contracts of the form ``Will event $X$ occur by date $Y$?''. The pre-event YES purchases at low prices, which form the empirical signature of insider activity in this literature, are the natural strategy on a deadline contract for a participant with private timing information. The present work is complementary to these studies: we use the patterns they identify to motivate feature design and to construct a labeled training set, but we constrain our detector to use only causal features observable while a market is active, and we extend the labeling score to the deadline contracts in which informed activity has been most clearly documented.

\subsection{Reflexivity, signal credibility, and proper-scoring evaluation}
\label{sec:related-reflexivity}

A separate strand of recent work treats prediction-market prices not as estimates to be evaluated for accuracy but as public signals that organize the behavior of voters, donors, journalists, and institutions. The classical reflexivity argument \citep{soros1987alchemy} is that publicly visible expectations can alter the world they purport to describe; in prediction markets, this dynamic operates through the conversion of dispersed private beliefs into a single salient focal point in the sense of \citet{schelling1960strategy}. \citet{manski2006interpreting} argued early that market-implied probabilities should not be read as Bayesian posteriors of well-defined events, and recent on-chain microstructure analyses \citep{tsang2026anatomy, tsang2026shocks, ng2026discovery} have shown that the social and informational properties of these signals diverge in measurable ways: the most cited platform during the 2024 U.S.~election cycle was simultaneously the least accurate by standard scoring metrics \citep{clinton2025prediction}.

\citet{nechepurenko2026focal} formalizes this divergence through the \emph{Signal Credibility Index} (SCI), a microstructure-grounded composite of the variance ratio, a two-sidedness diagnostic, and a trader-concentration adjustment, designed to predict when a price move will acquire behavioral traction. The variance-ratio component---comparing the variance of long-horizon returns to that of shorter-horizon returns---distinguishes durable, drift-like repricing from temporary order-flow pressure that quickly reverses. The two-sidedness component measures whether post-shock trading is consensual or contested. These diagnostics are exactly the kind of microstructural primitives needed to detect informed flow as well; SCI uses them to assess social authority, while ForesightFlow uses closely related quantities to assess informational provenance. The two frameworks are thus parallel applications of a shared microstructure toolkit to distinct downstream questions---is the price move socially consequential, and is it informationally provenanced.

A complementary line of work develops on-chain infrastructure for evaluating probabilistic forecasters using proper scoring rules \citep{brier1950verification, gneiting2007strictly} and the classical Murphy decomposition of the Brier score \citep{murphy1973decomposition}. \citet{nechepurenko2026foresight} introduce Foresight Arena, a permissionless on-chain benchmark that uses commit--reveal predictions on Polymarket markets, with scoring resolved through the Gnosis Conditional Token Framework. Their Alpha Score---defined as the Brier-score gap between an agent and the market consensus---is shown to decompose into a resolution gain and a reliability gap. The connection to the present work is structural: ILS (\Cref{sec:ils}) measures the fraction of resolution that has been accomplished by the market itself before the public information set has caught up, providing an explicit bridge between the proper-scoring framework and the informed-flow detection problem.

\subsection{Position of this work}
\label{sec:related-position}

\Cref{tab:related-positioning} summarizes the position of ForesightFlow relative to the immediately adjacent literature. The combination of classical microstructure measures with on-chain wallet features in a real-time detection framework, applied specifically to prediction-market data, has not, to our knowledge, been pursued in published work.

\begin{table}[t]
\centering
\caption{Position of ForesightFlow relative to adjacent prior work.}
\label{tab:related-positioning}
\small
\renewcommand{\arraystretch}{1.25}
\begin{tabularx}{\linewidth}{@{}p{4.6cm}Xccc@{}}
\toprule
\textbf{Work} & \textbf{Setting} & \textbf{Microstr.} & \textbf{On-chain} & \textbf{Real-time} \\
\midrule
\citet{easley1996liquidity}, PIN          & Equity, daily             & \checkmark & ---        & ---        \\
\citet{easley2012vpin}, VPIN              & Futures, intraday          & \checkmark & ---        & \checkmark \\
\citet{daian2020flash}, MEV               & DEX latency arbitrage      & ---        & \checkmark & \checkmark \\
\citet{imdea2025polymarket}               & Polymarket arbitrage       & ---        & \checkmark & ---        \\
\citet{mitts2026iran}                     & Polymarket informed screen & ---        & \checkmark & ---        \\
\citet{nechepurenko2026focal}             & Social authority of price  & \checkmark & \checkmark & \checkmark \\
\citet{nechepurenko2026foresight}         & On-chain forecasting bench & ---        & \checkmark & \checkmark \\
\citet{schoenegger2024large}              & LLM forecasting accuracy   & ---        & ---        & ---        \\
\midrule
\textbf{This work}                        & Prediction-market informed flow & \checkmark & \checkmark & \checkmark \\
\bottomrule
\end{tabularx}
\end{table}

\section{Data and Categorization}
\label{sec:data}

This section describes the data pipeline that ForesightFlow draws upon, the operational definitions of the three target categories, and the methodology for recovering the news timestamps that are central to the labeling procedure introduced in \Cref{sec:ils}.

\subsection{Polymarket primary data}
\label{sec:data-polymarket}

We use four complementary access paths to Polymarket data, each providing different parts of the picture. \Cref{tab:data-sources} summarizes them.

\begin{table}[t]
\centering
\caption{Data sources and their roles in the ForesightFlow pipeline.}
\label{tab:data-sources}
\small
\renewcommand{\arraystretch}{1.25}
\begin{tabularx}{\linewidth}{@{}llX@{}}
\toprule
\textbf{Source} & \textbf{Access} & \textbf{Use} \\
\midrule
Polymarket Gamma API     & REST, no auth          & Market metadata, tags, resolution criteria, end dates \\
Polymarket CLOB API      & REST + WebSocket       & Live and historical price/volume, full order book \\
Polymarket subgraph      & GraphQL via The Graph  & Full historical trade log per market with wallet addresses \\
UMA Optimistic Oracle    & On-chain + subgraph    & Resolution timestamps and proposer evidence URLs \\
Polygonscan API          & REST                   & Wallet-level on-chain context (first transaction, funding) \\
GDELT 2.0                & BigQuery               & First-public-mention timestamps for resolution-relevant events \\
\bottomrule
\end{tabularx}
\end{table}

The Gamma API supplies metadata about each market, including the original question text, structured tags assigned by Polymarket, the configured resolution date, and the resolution criteria. The CLOB API exposes both live and historical mid-price and volume data; for backfill we use the \texttt{prices-history} endpoint at one-minute resolution where available. The Polymarket subgraph, hosted on The Graph, exposes the full historical trade log: every fill, the addresses on both sides of the trade, the size, and the timestamp. Trade direction (buy versus sell of the YES outcome token) is explicit, eliminating the need for tick-rule classifiers used in classical microstructure work \citep{lee1991inferring}. UMA's Optimistic Oracle, used by Polymarket to resolve markets, exposes the resolution proposal itself and---critically for our purposes---the URL of the supporting evidence cited by the proposer. Polygonscan supplies wallet-level on-chain context.

\subsection{News timestamps}
\label{sec:data-news}

A central methodological challenge is the recovery, for each resolved market in our sample, of $\Tnews$: the timestamp of the first public mention of the event whose realization determined the market's resolution. Without $\Tnews$, the Information Leakage Score in \Cref{sec:ils} is undefined. We use a hierarchical procedure ordered by source authority.

The first authority is the UMA proposer evidence URL. When a Polymarket market resolves, the UMA proposer typically attaches a URL of the article, press release, or filing that they treat as authoritative for the resolution. The publication timestamp of this source is, by construction, the platform-recognized first public mention. We treat this as the gold standard for $\Tnews$ when available.

The second authority is the GDELT 2.0 Global Knowledge Graph \citep{leetaru2013gdelt}. GDELT continuously monitors news media in over one hundred languages and exposes a minute-resolution event index queryable via Google BigQuery. For markets without a usable proposer URL, we extract a small set of distinguishing keywords and named entities from the market question and resolution criteria, query the GDELT GKG for the earliest matching mention within a plausible window, and adopt that timestamp as $\Tnews$. The free BigQuery tier (1 TB of query data per month) is more than sufficient for the scope considered here.

The third tier, used only for the labeled validation set, is LLM-assisted matching against general web-search providers. This is computationally and financially expensive and is reserved for cases where the first two methods produce ambiguous or implausible results.

\begin{figure}[t]
\centering
\resizebox{\linewidth}{!}{%
\begin{tikzpicture}[
  every node/.style={font=\small},
  block/.style={rectangle, draw, rounded corners=2pt, align=center, minimum height=2.4em, minimum width=8em, inner sep=4pt},
  decision/.style={diamond, draw, aspect=2, align=center, inner sep=2pt},
  arrow/.style={-Latex, thick},
  node distance=0.7cm and 1.2cm
]

\node[block, fill=blue!8] (mkt) {Resolved market $M$};
\node[decision, right=of mkt, fill=yellow!15] (uma) {UMA evidence\\URL exists?};
\node[block, above right=0.4cm and 1.5cm of uma, fill=green!15] (gold) {Use article\\timestamp};
\node[decision, below right=0.4cm and 1.5cm of uma, fill=yellow!15] (gdelt) {GDELT GKG\\hit?};
\node[block, above right=0.4cm and 1.5cm of gdelt, fill=green!12] (gdeltuse) {Use earliest\\GDELT match};
\node[block, below right=0.4cm and 1.5cm of gdelt, fill=orange!18] (llm) {LLM-assisted\\matching};
\node[block, right=4.0cm of gdeltuse, fill=blue!8] (out) {$\Tnews$ for $M$};

\draw[arrow] (mkt) -- (uma);
\draw[arrow] (uma) -- node[above, sloped, font=\scriptsize] {yes} (gold);
\draw[arrow] (uma) -- node[below, sloped, font=\scriptsize] {no} (gdelt);
\draw[arrow] (gdelt) -- node[above, sloped, font=\scriptsize] {yes} (gdeltuse);
\draw[arrow] (gdelt) -- node[below, sloped, font=\scriptsize] {no} (llm);
\draw[arrow] (gold) -| (out);
\draw[arrow] (gdeltuse) -- (out);
\draw[arrow] (llm) -| (out);

\end{tikzpicture}%
}
\caption{Hierarchical recovery of the news timestamp $\Tnews$ for a resolved market. UMA proposer evidence is used when available; GDELT keyword matching is the default fallback; LLM-assisted matching is reserved for the labeled validation set.}
\label{fig:tnews-pipeline}
\end{figure}
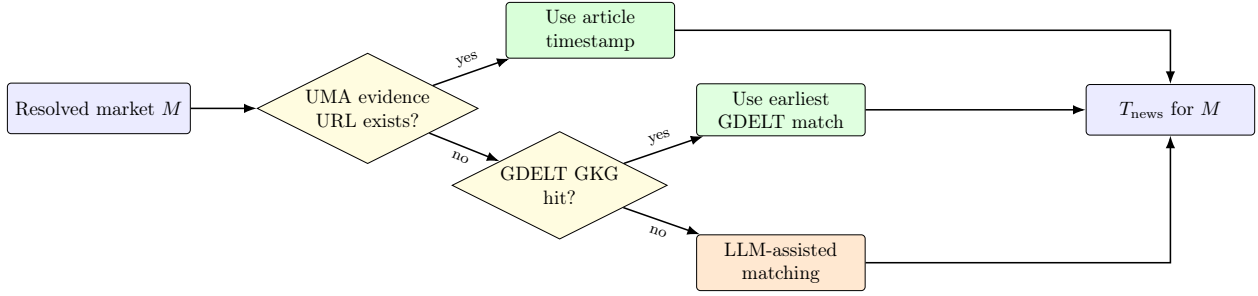

\Cref{fig:tnews-pipeline} summarizes the procedure.

\paragraph{Empirical caveat from the pilot.} The hierarchy above is the principled procedure; the empirical reality of the Polymarket corpus modifies it materially. Two regimes coexist on Polymarket: \emph{UMA-resolved markets}, where the resolution is performed via the UMA Optimistic Oracle and a proposer URL is on-chain, and \emph{admin-resolved markets}, where Polymarket itself pushes the resolution transaction without an on-chain evidence URL. Inspection of the corpus shows that UMA-resolved markets are dominated by automated-data-feed resolutions (Chainlink price oracles for crypto contracts; sports-result aggregator pages for game outcomes), where the evidence URL is a data feed rather than an article and the article-timestamp recovery in Tier~1 fails by construction. Admin-resolved markets, which include essentially all event-resolved political, regulatory, and corporate-disclosure contracts, have no on-chain evidence URL at all. The practical consequence is that for the population of markets where ILS is operationally most interesting, Tier~1 is empirically unavailable. The pilot in \Cref{sec:pilot} therefore proceeds with a resolution-anchored proxy, $\Tnews \equiv \Tres - 24\,\text{h}$, which we discuss as a methodological constraint rather than a design choice. GDELT-based Tier~2 recovery and LLM-assisted Tier~3 recovery remain the principled paths forward and are sequenced for subsequent revisions.

\subsection{Operational definitions of target categories}
\label{sec:data-categories}

Polymarket's native tag taxonomy is too coarse for our purposes: a single ``Politics'' or ``Business'' tag aggregates markets whose informational structure is fundamentally different. We define three high-priority categories operationally and augment the native taxonomy with an LLM-assisted fine-grained classifier whose output is then manually audited on a sample.

\begin{description}[leftmargin=1.4em, style=nextline]
\item[Military and geopolitical actions.] Markets resolving on \emph{specific} state actions whose date and content become public at the moment of announcement or execution. Includes military strikes, troop movements, diplomatic recognition, treaty signings, prisoner exchanges, sanction announcements, and embassy openings or closings. Excludes outcome of ongoing conflicts (``will war end by date $X$''), election outcomes, and generic geopolitical sentiment. Documented insider cases include the February 2026 strike on Iran, the January 2026 Venezuela operation, and a Maduro-capture market.

\item[Corporate proprietary disclosures.] Markets resolving on specific corporate events whose date or content is known to a narrow circle within the company prior to public announcement. Includes product launch dates, M\&A announcements, earnings beats or misses on specific metrics, executive hires or fires, regulatory filings, IP releases, and proprietary dataset publications such as Google's annual ``Year in Search'' rankings. Excludes stock-price levels and generic ``will company $X$ succeed'' markets. Documented insider cases include AlphaRaccoon's Year-in-Search trades and pre-launch trading on the OpenAI browser and Gemini 3.0 release date.

\item[Regulatory decisions.] Markets resolving on specific regulatory decisions with date-bounded resolution criteria. Includes FDA approvals, FCC rulings, SEC enforcement actions, central-bank rate decisions where the resolution is a concrete numerical level, court rulings, and antitrust decisions. Excludes generic ``will regulation $X$ happen this year'' markets and broad policy-direction predictions.
\end{description}

\textbf{Null-hypothesis controls.} Sports outcomes, weather forecasts, election polling levels, and cryptocurrency price levels serve as control categories. They are out of scope for detection; they are used solely to verify that the ILS distribution and microstructure features behave as expected under the null hypothesis of no informed trading.

\subsection{Sample}
\label{sec:data-sample}

The empirical sample covers Polymarket from October 2020 through April 2026, retrieved via the historical-backfill mode of the Gamma collector (querying \texttt{closed=true} with month-bucketed \texttt{end\_date} ranges). After deduplication, the database contains 911{,}237 markets, of which 865{,}725 are resolved. \Cref{tab:sample-distribution} summarizes the distribution across the three target categories and the residual control category, both at the full-sample level and at a high-volume cutoff that we adopt for trade-level analysis.

\begin{table}[t]
\centering
\caption{Sample distribution after the 2020--2026 historical backfill. Categories are assigned by the keyword-based taxonomy classifier described in \Cref{sec:data-categories}; native Polymarket tags were found to be non-functional for closed markets and are not used. The high-volume cutoff at \$50{,}000 is empirically motivated: below this threshold the Polymarket subgraph typically contains no CLOB trade history (markets resolve via automated oracles or direct data feeds without on-book trading).}
\label{tab:sample-distribution}
\small
\renewcommand{\arraystretch}{1.25}
\begin{tabularx}{\linewidth}{@{}lXXXX@{}}
\toprule
\textbf{Category} & \textbf{Total} & \textbf{Resolved} & \textbf{Resolved, $\geq$\$50K vol} & \textbf{\%} \\
\midrule
military / geopolitics    & 47{,}580   & 44{,}436   & 3{,}970 & 9.0\% \\
regulatory decisions      & 71{,}588   & 65{,}542   & 5{,}582 & 8.5\% \\
corporate disclosures     & 20{,}645   & 17{,}425   & 1{,}711 & 9.8\% \\
other (control)           & 771{,}424  & 738{,}322  & 88{,}656 & 12.0\% \\
\midrule
\textbf{Total}            & 911{,}237  & 865{,}725  & 99{,}919 &       \\
\bottomrule
\end{tabularx}
\end{table}

Two empirical observations from the inventory deserve emphasis as they shape the methodology that follows.

\paragraph{Volume bifurcation and CLOB coverage.} The Polymarket subgraph indexes only orders matched on the central limit order book. A substantial fraction of low-volume resolved markets---routine sports outcomes, daily weather forecasts, hourly cryptocurrency price ticks---resolve through automated oracle feeds (e.g., Chainlink, sport-specific data providers) without ever generating CLOB trades. Empirically, markets with volume below approximately \$50{,}000 consistently return zero trades from the subgraph, whereas markets above this threshold reliably have full trade histories. We therefore treat the \$50{,}000 threshold as a hard prerequisite for any analysis that depends on order-flow features, including all microstructure and wallet-level signals (\Cref{sec:micro}). The threshold is not a free parameter that can be tuned; it reflects an infrastructure boundary of the Polymarket platform itself.

\paragraph{Structural break around 2024.} The growth of Polymarket between 2020 and 2026 is not uniform: total volume grew by more than three orders of magnitude over the period, with the inflection point centered on the 2024 U.S. presidential election cycle. The qualitative consequence is that markets resolved before mid-2024 are substantively different objects from markets resolved after: same outcome categories, but different liquidity, different participant base, different venue maturity. Statistical analyses pooled across the full window will conflate these regimes and should be reported with explicit time-stratified breakdowns. We adopt this convention throughout the planned empirical companion.

\subsection{Validation set: documented insider cases}
\label{sec:data-validation}

A central methodological constraint of any informed-flow detection framework is the scarcity of unambiguously labelled positive examples. To address this, we construct a curated validation set, the \emph{ForesightFlow Insider Cases} (FFIC) inventory, drawn from publicly reported episodes between 2024 and early 2026 in which press coverage, academic discussion, or platform-side disclosure has documented or strongly implied informed positioning ahead of resolution. Each case maps to one or more concrete Polymarket market identifiers and is anchored to a public reference. \Cref{tab:ffic} summarizes the inventory.

\begin{table}[t]
\centering
\caption{ForesightFlow Insider Cases (FFIC) inventory. Each case corresponds to a publicly reported episode of suspected informed trading on Polymarket. The volume column reports the maximum traded notional across the markets associated with the case; the indexer column flags cases for which The Graph subgraph currently does not return trade data due to indexer-capacity limits (an infrastructure constraint discussed in \Cref{sec:micro-power}).}
\label{tab:ffic}
\small
\renewcommand{\arraystretch}{1.25}
\begin{tabularx}{\linewidth}{@{}p{4.8cm}lXll@{}}
\toprule
\textbf{Case} & \textbf{Date} & \textbf{Category} & \textbf{Volume} & \textbf{Indexer} \\
\midrule
2024 U.S. presidential election    & 2024-11   & politics / military$^{\dagger}$ & \$2.57B & limited \\
Iran strike on Israel              & 2024-10-01 & military / geopolitics & \$148K & ok \\
Romanian election (Ciuc\u{a})      & 2025-05   & regulatory / political & \$326M & ok \\
Maduro / Venezuela cluster         & 2026-01--02 & military / geopolitics & \$89M & ok \\
2026 U.S.--Iran conflict cluster   & 2026-02--04 & military / geopolitics & \$269M & limited \\
Bitcoin ETF SEC approval           & 2024-01-10 & regulatory decision & \$12.6M & ok \\
Google Year-in-Search 2025         & 2025-12   & corporate disclosure & \$3M & ok \\
FTX / SBF cluster                  & 2024--2025 & regulatory decision & \$8.2M & ok \\
\bottomrule
\end{tabularx}

\smallskip
\footnotesize
$^{\dagger}$ The 2024 U.S.\ presidential election does not fit cleanly into a single target category; it is treated as a politics-adjacent military / geopolitics case for the purposes of this validation set, with explicit acknowledgement of the categorization stretch.
\end{table}

The inventory exhibits two structural patterns that are themselves informative for the methodology.

\paragraph{Volume is not a proxy for documented informed activity.} The October 2024 Iran strike market resolved at a peak volume of approximately \$148{,}000---two to three orders of magnitude smaller than the 2026 U.S.--Iran cluster covering structurally analogous events---yet attracted more public scrutiny and is among the most concretely documented cases. This empirically refutes the natural assumption that informed trading should be sought primarily in the highest-volume markets, and motivates a category-conditional approach (\Cref{sec:micro-vector}) rather than a volume-weighted one.

\paragraph{The Graph indexer constraint at extreme volume.} For the largest markets in the inventory---the 2024 U.S.\ presidential top-line markets, the largest 2026 Iran-conflict markets---the public Polymarket subgraph indexer returns errors of the form \texttt{bad indexers}, indicating that no indexer in The Graph network has fully indexed these contracts due to their size. This is a hard infrastructure limit on what can be observed via the standard subgraph path; recovering trade-level data for these markets requires either an alternate indexing path (e.g., a self-hosted subgraph instance or direct event-log decoding via Polygon JSON-RPC) or acceptance that the very largest markets are excluded from microstructure-level analysis. We adopt the latter for the present version of the work and document it explicitly as a limitation (\Cref{sec:limitations}).

The FFIC inventory will be released as a versioned artifact alongside this paper, with structured per-case manifests including market identifiers, event timelines, and references to public sources. The inventory is intended as a community resource: future work on informed-trading detection in prediction markets can be directly compared against ForesightFlow on the same labelled cases, addressing one of the persistent reproducibility weaknesses of the on-chain forensics literature.

\section{Information Leakage Score}
\label{sec:ils}

The Information Leakage Score (ILS) is the central label generator of the ForesightFlow framework. It quantifies, for any resolved binary market with an identifiable news event, the fraction of the total information move that occurred before the news. Below we give the formal definition, identify the resolution-typology and scope conditions under which the score is interpretable, discuss its scaling and edge-case behaviour, connect it to the proper-scoring-rule literature via the Murphy decomposition, define the auxiliary metrics that complete the labeling pipeline, and describe how the labels feed into the real-time detector. \Cref{tab:notation} lists the notation used throughout this and the following sections.

\begin{table}[h]
\centering
\caption{Notation used in the score definition and its extension. Timestamps are UTC unless noted; prices are mid-quotes on the YES outcome token, $p \in [0,1]$.}
\label{tab:notation}
\small
\renewcommand{\arraystretch}{1.2}
\begin{tabularx}{\linewidth}{@{}lX@{}}
\toprule
\textbf{Symbol} & \textbf{Meaning} \\
\midrule
$\Topen$         & Market creation timestamp (first trade) \\
$\Tnews$         & First public mention of resolution-relevant information \\
$\Tres$          & Formal resolution timestamp (UMA Optimistic Oracle or Polymarket admin) \\
$\Tevent$        & Public observation of the underlying event (deadline markets, \Cref{sec:extension}) \\
$\Tdead$         & Deadline-expiration timestamp for deadline-resolved markets \\
$p(t)$           & Mid-price of the YES outcome token at time $t$ \\
$p_{\Tres}$      & Binary resolution outcome ($\{0, 1\}$) \\
$\theta_t$       & Bayesian-baseline belief at time $t$ for deadline markets \\
$\Delta_{\text{pre}}$    & Pre-news price drift, $p(\Tnews) - p(\Topen)$ \\
$\Delta_{\text{total}}$  & Total information move, $p_{\Tres} - p(\Topen)$ \\
$\ILS(M)$        & Information Leakage Score for market $M$ (event-resolved) \\
$\ILS^{\text{dl}}(M)$    & Deadline-ILS extension (\Cref{sec:extension}) \\
$\varepsilon$    & Trivial-resolution threshold; default $\varepsilon = 0.05$ \\
$S(\tau \mid \Topen)$    & Survival function of time-to-event conditional on info at $\Topen$ \\
\bottomrule
\end{tabularx}
\end{table}

\subsection{Definition}
\label{sec:ils-def}

Consider a resolved binary market $M$ with three associated timestamps:
\begin{itemize}[leftmargin=1.4em, itemsep=0.1em]
\item $\Topen$ — market creation (first trade);
\item $\Tnews$ — first public mention of resolution-relevant information, recovered as in \Cref{sec:data-news};
\item $\Tres$ — UMA Optimistic Oracle resolution.
\end{itemize}
Let $p(t) \in [0, 1]$ denote the mid-price of the YES outcome token at time $t$, and let $p_{\Tres} \in \{0, 1\}$ be the binary resolution.

\begin{definition}[Information Leakage Score]
\label{def:ils}
The pre-news drift, total information move, and Information Leakage Score for market $M$ are
\[
\Delta_{\mathrm{pre}} \;=\; p(\Tnews) - p(\Topen),
\qquad
\Delta_{\mathrm{total}} \;=\; p_{\Tres} - p(\Topen),
\qquad
\ILS(M) \;=\; \frac{\Delta_{\mathrm{pre}}}{\Delta_{\mathrm{total}}},
\]
where the score is defined whenever $|\Delta_{\mathrm{total}}| > \varepsilon$ for some small threshold $\varepsilon > 0$.
\end{definition}

The threshold $\varepsilon$ rules out markets whose price barely moved between opening and resolution; the score is uninformative in that regime. We use $\varepsilon = 0.05$ in the empirical work, corresponding to a 5-cent total move.

\subsection{Interpretation}
\label{sec:ils-interp}

\Cref{tab:ils-interp} summarizes the regimes of the ILS. The natural reading is that $\ILS(M) = 1$ corresponds to fully priced-in information before the news event---perfect anticipatory pricing---while $\ILS(M) = 0$ corresponds to a market that did not move at all in the pre-news window and reacted only to the public announcement.

\begin{table}[t]
\centering
\caption{Interpretation of ILS regimes.}
\label{tab:ils-interp}
\small
\renewcommand{\arraystretch}{1.25}
\begin{tabularx}{\linewidth}{@{}lX@{}}
\toprule
\textbf{Regime} & \textbf{Interpretation} \\
\midrule
$\ILS \approx 1$         & Full information move priced before public news (strong leakage). \\
$\ILS \approx 0$         & Pre-news price flat; market reacted to public news as expected. \\
$\ILS \in (0, 1)$        & Partial leakage; some informed flow before news, completed afterwards. \\
$\ILS > 1$               & Pre-news overshoot of the eventual outcome (overreaction or correctly aimed speculation). \\
$\ILS < 0$               & Pre-news price moved against the eventual outcome (counter-evidence). \\
\bottomrule
\end{tabularx}
\end{table}

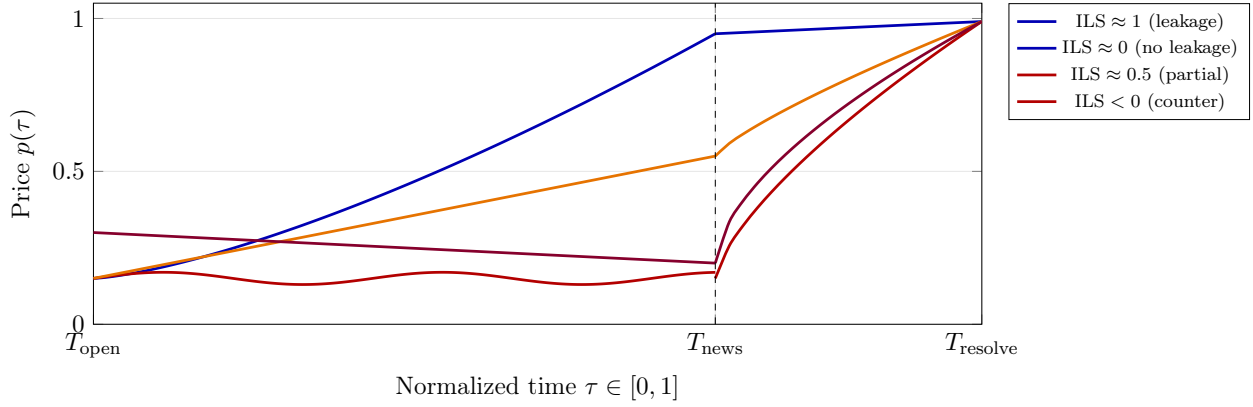
\begin{figure}[t]
\centering
\resizebox{\linewidth}{!}{%
\begin{tikzpicture}
\begin{axis}[
  width=0.92\linewidth, height=6.5cm,
  xlabel={Normalized time $\tau \in [0, 1]$},
  ylabel={Price $p(\tau)$},
  xmin=0, xmax=1, ymin=0, ymax=1.05,
  xtick={0, 0.7, 1}, xticklabels={$\Topen$, $\Tnews$, $\Tres$},
  ytick={0, 0.5, 1},
  legend pos=outer north east,
  legend style={font=\scriptsize},
  grid=major, grid style={very thin, gray!20},
  every axis plot/.append style={very thick}
]

\addplot[blue!70!black, smooth, samples=80, domain=0:0.7] {0.15 + (0.95-0.15)*(x/0.7)^1.4};
\addplot[blue!70!black, smooth, samples=20, domain=0.7:1] {0.95 + 0.04*(x-0.7)/0.3};
\addlegendentry{$\ILS \approx 1$ (leakage)}

\addplot[red!70!black, smooth, samples=80, domain=0:0.7] {0.15 + 0.02*sin(deg(20*x))};
\addplot[red!70!black, smooth, samples=20, domain=0.7:1] {0.15 + (0.99-0.15)*((x-0.7)/0.3)^0.7};
\addlegendentry{$\ILS \approx 0$ (no leakage)}

\addplot[orange!90!black, smooth, samples=80, domain=0:0.7] {0.15 + (0.55-0.15)*(x/0.7)^1.0};
\addplot[orange!90!black, smooth, samples=20, domain=0.7:1] {0.55 + (0.99-0.55)*((x-0.7)/0.3)^0.8};
\addlegendentry{$\ILS \approx 0.5$ (partial)}

\addplot[purple!70!black, smooth, samples=80, domain=0:0.7] {0.30 - 0.10*(x/0.7)};
\addplot[purple!70!black, smooth, samples=20, domain=0.7:1] {0.20 + (0.99-0.20)*((x-0.7)/0.3)^0.6};
\addlegendentry{$\ILS < 0$ (counter)}

\addplot[black, dashed, very thin, forget plot] coordinates {(0.7, 0) (0.7, 1.05)};

\end{axis}
\end{tikzpicture}%
}
\caption{Four characteristic price trajectories, normalized to $\tau \in [0, 1]$, illustrating different ILS regimes for a market that resolves YES. The vertical dashed line marks $\Tnews$. Trajectories with steep ascent before the dashed line correspond to high-leakage regimes; trajectories that move predominantly after the line correspond to no-leakage or low-leakage regimes.}
\label{fig:ils-regimes}
\end{figure}

\Cref{fig:ils-regimes} visualizes four typical price trajectories on a normalized time axis, illustrating how the same definition recovers materially different leakage regimes from the shape of $p(t)$.

\subsection{Resolution typology and the scope of ILS}
\label{sec:ils-typology}

ILS as defined in \Cref{sec:ils-def} presupposes a well-defined news timestamp $\Tnews$. Empirical examination of the Polymarket corpus shows that this presupposition is non-trivial: a substantial fraction of markets resolve in a manner that makes $\Tnews$ either undefined or trivially equal to $\Tres$. We classify resolved markets into three operational types.

\paragraph{Event-resolved markets.} The market resolves when a specific observable event occurs, with a clear timestamp. Examples: an electoral outcome, a regulatory decision, the public release of a corporate filing, the start of a military operation. For these markets, $\Tnews$ is well-defined and ILS is meaningful.

\paragraph{Deadline-resolved markets.} The market resolves NO when a specified event \emph{fails to occur} by a fixed deadline (``Will Iran strike Israel by Friday?'' resolved NO; ``Tesla launches FSD by October 31?'' resolved NO). For these markets, the canonical ``news event'' is the absence of an event, and there is no informative public moment that the market could have anticipated. The concept of $\Tnews$ does not apply, and ILS values computed on these markets reflect price-convergence noise rather than informed flow.

\paragraph{Surprise-resolved markets.} The market resolves opposite to its consensus probability (price near 99\% throughout, resolves NO; or vice versa). For these markets, the price did not move toward resolution at all, and ILS is structurally near zero --- but the interpretation is the opposite of an efficient market: it is consistent with the absence of any informed positioning, not with its presence.

A keyword-based heuristic classifier on 11{,}200 resolved markets in the three target categories above the \$50K volume cutoff produces the distribution in \Cref{tab:resolution-typology}.

\begin{table}[t]
\centering
\caption{Resolution-type distribution across 11{,}200 resolved markets in target categories with volume $\geq \$50$K. The 100\% NO rate among deadline-resolved markets is structural --- the classifier identifies markets whose resolution criterion is the non-occurrence of an event. The unclassifiable bucket comprises sports outcomes and metric-count markets where the heuristic is conservative; a future taxonomy refinement is expected to recover a portion of these as event-resolved.}
\label{tab:resolution-typology}
\small
\renewcommand{\arraystretch}{1.25}
\begin{tabularx}{\linewidth}{@{}lXrrr@{}}
\toprule
\textbf{Type} & \textbf{Description} & \textbf{N} & \textbf{\%} & \textbf{YES rate} \\
\midrule
event-resolved      & Specific observable event triggers resolution                  & 1{,}145  & 10.2\%  & 29.1\% \\
deadline-resolved   & Resolves NO if event fails to occur by fixed deadline          & 1{,}224  & 10.9\%  & 0.0\%  \\
unclassifiable      & Sports, count markets, ambiguous patterns                      & 8{,}831  & 78.8\%  & 31.5\% \\
\midrule
\textbf{Total}      &                                                                  & 11{,}200 &         &        \\
\bottomrule
\end{tabularx}
\end{table}

The 0\% YES rate among deadline-resolved markets is a structural validation of the typology: the classifier independently identifies markets whose resolution criterion is, by construction, the failure of an event to occur. ILS computation in the rest of this work is restricted to event-resolved markets. Deadline-resolved and unclassifiable markets remain in the corpus but are excluded from the ILS-based analytical pipeline; they may serve as control populations or as targets for distinct methodologies in future work.

\subsection{Edge cases and scope conditions}
\label{sec:ils-scope}

The compactness of the ILS definition obscures three regimes in which raw ILS values are not interpretable as evidence of informed flow. We make these explicit because, as the pilot study reported in \Cref{sec:pilot} demonstrates, all three appear in non-trivial proportions in real Polymarket data, and conflating them with genuine leakage produces spurious positive findings.

\paragraph{Edge-effect regime: $|p_{\Tres} - p_{\Topen}|$ is small.} When a market opens at a near-consensus price (e.g., $p_{\Topen} = 0.94$ for an outcome that resolves YES), the denominator $\Delta_{\text{total}} = p_{\Tres} - p_{\Topen}$ is small by construction, and any small pre-news drift is amplified into a near-unit ILS value. Concretely, a market that opens at $0.94$ and reaches $0.996$ before $\Tnews$ has $\ILS = 0.056 / 0.060 \approx 0.93$, despite an absolute pre-news price move of only six percentage points and an aggregate market consensus that was already strong from inception. The high ILS in this regime reflects the geometry of the formula, not any informational asymmetry. We adopt the operational rule that ILS is reported as informative only for markets satisfying $|p_{\Topen} - 0.5| \leq 0.4$, equivalently $p_{\Topen} \in [0.1, 0.9]$, and flag markets outside this range as edge-effect-dominated. Multi-window variants (\Cref{sec:ils-windows}) inherit the same scope condition applied to $p(\Tnews - w)$.

\paragraph{Trivial-resolution regime: $|\Delta_{\text{total}}| < \varepsilon$.} When a market never moves materially over its lifetime---typically because its outcome was never genuinely uncertain---ILS is undefined. We treat this as the explicit threshold $|\Delta_{\text{total}}| < \varepsilon$ with $\varepsilon = 0.05$, in which case ILS is reported as missing rather than as zero. A market in this regime contains no informational signal of either direction.

\paragraph{Anchor-sensitivity regime.} ILS is a ratio anchored at a discrete time $\Tnews$, and the recovered value depends on the choice of anchor. For markets with a well-recovered article-derived $\Tnews$, this is unambiguous. For markets where $\Tnews$ is approximated by a proxy (e.g., $\Tres$ minus a fixed offset), ILS values that change qualitatively with the offset choice are not robust signals. We make this concrete in \Cref{sec:ils-windows} by requiring that any high-ILS claim be accompanied by a multi-window robustness check; if the sign or rough magnitude of ILS is not preserved across at least two distinct anchor specifications, the value is treated as proxy-artefact rather than as informational.

These three scope conditions are restrictive: they exclude a substantial fraction of resolved Polymarket markets from interpretable ILS analysis. We view this as a feature of the methodology rather than a defect. The alternative---reporting raw ILS values across the full corpus and treating high-magnitude values as evidence of leakage---produces results that, as the pilot study shows, are not statistically distinguishable from the null distribution.

\subsection{Multi-window variants}
\label{sec:ils-windows}

The single-point definition above collapses the entire pre-news interval to one number. In practice, the temporal profile of price drift before $\Tnews$ is itself informative: a leakage that accumulates uniformly over weeks differs operationally from one concentrated in the final hours. We define multi-window ILS variants

\[
\ILS_{w}(M) \;=\; \frac{p(\Tnews) - p(\Tnews - w)}{p_{\Tres} - p(\Tnews - w)}
\]

for windows $w \in \{30\,\text{min}, 2\,\text{h}, 6\,\text{h}, 24\,\text{h}, 7\,\text{d}\}$. The resulting vector $(\ILS_{30\text{min}}, \ldots, \ILS_{7\text{d}})$ characterizes the timing profile of the leakage. For real-time detection, the short-window variants are the operationally relevant ones, as they capture the regime in which informed trading is most likely to be ongoing during the detection window.

\subsection{Connection to the Murphy decomposition}
\label{sec:ils-murphy}

The classical Murphy decomposition \citep{murphy1973decomposition} expresses the Brier score of a probabilistic forecaster as the sum of three terms:
\[
B \;=\; \mathrm{UNC} \;+\; \mathrm{REL} \;-\; \mathrm{RES},
\]
where $\mathrm{UNC} = \bar{o}(1 - \bar{o})$ is the irreducible outcome uncertainty, $\mathrm{REL}$ is calibration error, and $\mathrm{RES}$ is the resolution component, capturing the squared gap between bin-conditional and unconditional outcome frequencies. Lower Brier requires higher resolution: a forecaster who never deviates from the base rate has $\mathrm{RES} = 0$.

The market-implied probability $p(t)$ can itself be viewed as a sequence of forecasts. Define $\mathrm{RES}(t)$ as the resolution component computed using market predictions up to time $t$. Then $\mathrm{RES}(\Tres)$ is the total resolution accomplished by the market over the entire market lifetime, and $\mathrm{RES}(\Tnews)$ is the portion already accomplished before public news.

\begin{remark}[ILS as front-loaded resolution]
\label{rem:ils-as-resolution}
Under regularity conditions on the price process---specifically, that pre-news price moves are unbiased estimators of the conditional outcome probability and that bin-conditional outcome frequencies converge to bin means---ILS is approximately the share of the market's total resolution component that was accumulated before $\Tnews$:
\[
\ILS(M) \;\approx\; \frac{\mathrm{RES}(\Tnews)}{\mathrm{RES}(\Tres)}.
\]
A high-ILS market is one whose discriminative power was front-loaded: it had already separated outcome from base rate before any public information justified that separation. Under the null hypothesis of an efficient market with no private information, $\mathrm{RES}(\Tnews) \approx 0$ and ILS $\approx 0$, with all resolution accruing through the post-news price reaction.
\end{remark}

\begin{figure}[t]
\centering
\resizebox{0.9\linewidth}{!}{%
\begin{tikzpicture}
\begin{axis}[
  ybar stacked,
  width=12cm, height=5.5cm,
  bar width=22pt,
  ymin=0, ymax=0.30,
  xtick=data,
  symbolic x coords={Null (no leakage), Partial leakage, Full leakage},
  xticklabel style={font=\small},
  ylabel={Brier components},
  ylabel style={font=\small},
  legend style={at={(1.02,1)}, anchor=north west, font=\scriptsize, draw=none},
  legend cell align=left,
  every axis plot/.append style={thick},
  enlarge x limits=0.25,
  axis lines=left,
  tick style={black},
  major grid style={very thin, gray!25},
  ymajorgrids=true,
]
\addplot[fill=blue!20, draw=blue!50!black] coordinates
  {(Null (no leakage), 0.250) (Partial leakage, 0.250) (Full leakage, 0.250)};
\addplot[fill=red!25, draw=red!60!black] coordinates
  {(Null (no leakage), 0.005) (Partial leakage, 0.010) (Full leakage, 0.012)};
\addplot[fill=green!28, draw=green!50!black] coordinates
  {(Null (no leakage), 0.010) (Partial leakage, 0.060) (Full leakage, 0.108)};
\legend{UNC ($\bar{o}(1{-}\bar{o})$), REL (calibration error), RES (resolution -- pre-news)}
\end{axis}
\end{tikzpicture}%
}
\caption{Murphy decomposition $B = \mathrm{UNC} + \mathrm{REL} - \mathrm{RES}$ at the news timestamp $\Tnews$, evaluated for three idealized regimes of the same set of markets. The uncertainty term $\mathrm{UNC}$ is fixed by the marginal outcome distribution and is identical across scenarios. The reliability error $\mathrm{REL}$ is small for any approximately-calibrated forecaster. The resolution component $\mathrm{RES}(\Tnews)$ is the only component that varies with the degree of pre-news leakage; \Cref{rem:ils-as-resolution} formalizes ILS as the share of total resolution that has accumulated by $\Tnews$. Numerical values are illustrative.}
\label{fig:murphy-decomp}
\end{figure}
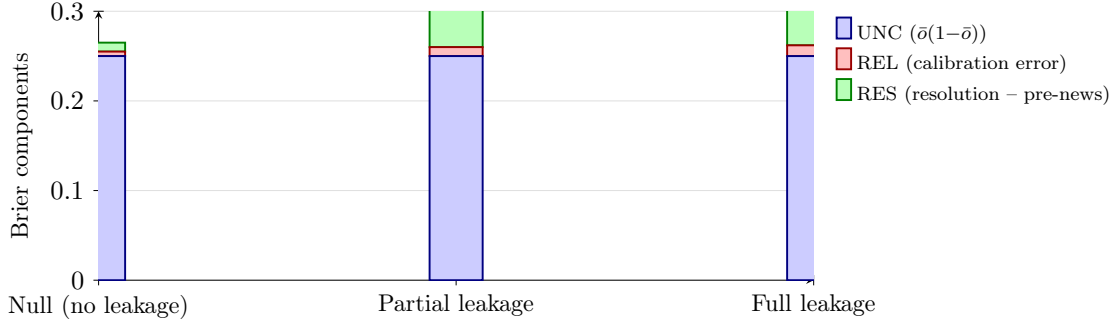

This connection is more than aesthetic. It places ILS on the same theoretical footing as the proper-scoring-rule literature, links the present work to the on-chain forecasting framework of \citet{nechepurenko2026foresight}, and clarifies the sense in which informed flow is detectable: it is the component of market discriminative power that arrives ahead of its public-information justification. The full Murphy decomposition can also be applied diagnostically to the agent-side detector: a real-time detector is, formally, a probabilistic classifier mapping observable features to a binary informed-flow label, and its calibration and discriminative power can be evaluated using the same UNC/REL/RES partition. We return to this in \Cref{sec:micro}.

\subsection{Auxiliary metrics}
\label{sec:ils-aux}

ILS by itself describes only the price trajectory; informed trading also leaves signatures in volume, in the size and timing of individual trades, and in the identities of the wallets responsible for the largest winning trades. We define four auxiliary metrics that, together with ILS, constitute the full label vector for each resolved market.

\paragraph{Pre-news volume share.}
\[
V_{\mathrm{pre}}(M) \;=\; \frac{\sum_{t < \Tnews} v(t)}{\sum_{t \leq \Tres} v(t)},
\]
where $v(t)$ is the volume traded at time $t$. High $V_{\mathrm{pre}}$ combined with high $\ILS$ is the prototypical leakage signature.

\paragraph{Pre-news price jump.} The maximum single-trade price impact in $[\Topen, \Tnews]$. This isolates one-shot informed trades, which large enough flow can produce, against the alternative of incremental drift.

\paragraph{Wallet concentration index (HHI).} For the top-$k$ winning trades in $M$, the Herfindahl--Hirschman index of trade-size shares,
\[
\mathrm{HHI}(M, k) \;=\; \sum_{i=1}^{k} s_i^2, \qquad s_i \;=\; \frac{q_i}{\sum_{j \leq k} q_j},
\]
where $q_i$ is the size of the $i$-th winning trade. We use $k = 10$.

\paragraph{Time-to-news distribution.} For each of the top-$k$ winning trades, the gap $\Tnews - t_i$ between the trade timestamp $t_i$ and the news timestamp. A distribution heavily concentrated near zero (trades clustered just before news) is a strong leakage signature; a distribution roughly uniform over $[\Topen, \Tnews]$ is consistent with an innocent informed-by-public-context interpretation.

\subsection{Wallet novelty score}
\label{sec:wn}

The on-chain transparency of Polymarket allows us to define wallet-level features that have no analog in traditional venues. The composite Wallet Novelty Score is, for a wallet $w$ at trade time $t$,
\begin{align*}
\WN(w, t) \;=\;\;\; & \alpha_1 \, \one\!\bigl[\mathrm{age}(w, t) < 48\,\text{h}\bigr]
                     + \alpha_2 \, \one\!\bigl[|\text{markets}(w, < t)| < 3\bigr] \\
                  & {} + \alpha_3 \, c_{\text{fund}}(w)
                     + \alpha_4 \, \one\!\bigl[t > \Tres - 2\,\text{h}\bigr],
\end{align*}
where $\mathrm{age}(w, t)$ is the time since the first on-chain transaction of $w$, $\text{markets}(w, < t)$ is the set of distinct Polymarket markets in which $w$ has previously traded, $c_{\text{fund}}(w) \in [0, 1]$ is a funding-source-concentration measure derived from the wallet's historical USDC inflows, and the final term flags entries within two hours of resolution. The weights $\alpha_i$ are fitted on the labeled training set; we do not commit to specific values in this draft.

\subsection{Label aggregation and detector input}
\label{sec:ils-detector}

For each resolved market $M$ in the training set that satisfies the scope conditions of \Cref{sec:ils-scope} and the resolution-typology restriction of \Cref{sec:ils-typology}, we record the label vector
\[
y(M) \;=\; \bigl(\ILS(M),\; \ILS_{30\text{min}}(M),\; \ILS_{2\text{h}}(M),\; V_{\mathrm{pre}}(M),\; \mathrm{HHI}(M, 10),\; \overline{\WN}(M)\bigr),
\]
where $\overline{\WN}(M)$ is the mean Wallet Novelty Score of the top-10 winning trades in $M$. For supervised training of the real-time detector, we collapse this vector into a binary label
\[
y_{\mathrm{bin}}(M) \;=\; \one\!\bigl[\ILS(M) \geq \theta_1 \;\wedge\; V_{\mathrm{pre}}(M) \geq \theta_2 \;\wedge\; \overline{\WN}(M) \geq \theta_3\bigr],
\]
with thresholds $(\theta_1, \theta_2, \theta_3)$ chosen on the validation set against case-study ground truth. The pilot of \Cref{sec:pilot} demonstrates that ILS alone is insufficient to materialize a clean labelled set under proxy-derived $\Tnews$; the joint thresholding above, together with the resolution typology and the scope conditions, is the operational form of that lesson. We do not commit to specific threshold values in this draft; they will be fitted once the deadline-ILS extension of \Cref{sec:extension} is implemented and the FFIC inventory becomes empirically addressable.

\section{Microstructure Signatures and Detector Architecture}
\label{sec:micro}

This section describes the real-time feature pipeline. The objective at inference time is to compute, for each active market in a target category, a calibrated probability that informed flow is currently moving the price---without using any feature that is observable only at or after resolution.

\subsection{From classical microstructure to discrete binary markets}
\label{sec:micro-adapt}

Three properties of the Polymarket setting force a non-trivial adaptation of classical microstructure measures.

First, prices are bounded in $[0, 1]$. The ratio-based formulations of price impact familiar from equity markets do not directly apply; we work in absolute price differences and normalize by the time-to-resolution where appropriate.

Second, trade direction is explicit on-chain. The classical PIN/VPIN literature relies on the Lee--Ready tick rule \citep{lee1991inferring} or analogous heuristics to classify trades as buyer- or seller-initiated; the Polymarket subgraph records this directly. Microstructure measures that depend on classification accuracy are correspondingly more reliable.

Third, the market is hybrid in execution: order matching is off-chain for latency reasons, but settlement is on-chain. This means that the classical assumption of continuous-quote dynamics is approximately satisfied within a single venue, but the chain-level observation of settled trades introduces a discretization at the on-chain settlement granularity (Polygon block time, currently approximately two seconds).

We adapt three families of measures.

\paragraph{Order imbalance.} For a rolling window of length $\Delta$ ending at $t$,
\[
\mathrm{OI}(t, \Delta) \;=\; \frac{V_{\mathrm{buy}}(t-\Delta, t) - V_{\mathrm{sell}}(t-\Delta, t)}{V_{\mathrm{buy}}(t-\Delta, t) + V_{\mathrm{sell}}(t-\Delta, t)} \;\in\; [-1, 1],
\]
where $V_{\mathrm{buy}}$ and $V_{\mathrm{sell}}$ are the volumes of YES-token buys and sells respectively. We compute $\mathrm{OI}$ at $\Delta \in \{5\,\text{min}, 15\,\text{min}, 1\,\text{h}\}$. Sustained high $|\mathrm{OI}|$ in a thin market with no public news is a microstructural informed-flow signature.

\paragraph{VPIN-style toxicity.} Following \citet{easley2012vpin}, we partition the cumulative volume of $M$ into volume buckets of size $V$, and within each bucket compute the absolute imbalance
\[
\mathrm{Tox}_n \;=\; \frac{|V_{\mathrm{buy}}^{(n)} - V_{\mathrm{sell}}^{(n)}|}{V}.
\]
The VPIN at time $t$ is the trailing average of $\mathrm{Tox}_n$ over the last $N$ buckets. The bucket size $V$ is tuned per category to match the typical market liquidity.

\paragraph{Kyle's lambda.} Within a rolling window we estimate
\[
\Delta p \;=\; \lambda \cdot \mathrm{NetFlow} + \eta,
\]
by ordinary least squares, where $\Delta p$ is the per-trade price change, $\mathrm{NetFlow}$ is the signed trade size, and $\eta$ is residual. Higher estimated $\lambda$ corresponds to higher per-unit price impact and is consistent with informed flow.

\paragraph{Variance ratio.} A complementary diagnostic, drawn from \citet{nechepurenko2026focal}, is the variance ratio comparing the variance of long-window log-returns to a multiple of the variance of short-window returns:
\[
\mathrm{VR}(k)(t) \;=\; \frac{\mathrm{Var}\bigl(r^{(k\Delta)}_{\le t}\bigr)}{k \cdot \mathrm{Var}\bigl(r^{(\Delta)}_{\le t}\bigr)}.
\]
Under a random walk, $\mathrm{VR}(k) = 1$. Values above unity indicate drift---the price is reproducing its short-horizon moves over longer horizons, consistent with persistent informed pressure---while values below unity indicate mean-reversion, consistent with temporary order-flow pressure that subsequently corrects. We use $k = 6$ with $\Delta = 5$~minutes, matching the convention of \citet{tsang2026shocks}.

\paragraph{Two-sidedness.} Following \citet{nechepurenko2026focal}, the two-sidedness index over a rolling window is
\[
\mathrm{TS}(t) \;=\; 1 - \frac{|V_{\mathrm{buy}}(t) - V_{\mathrm{sell}}(t)|}{V_{\mathrm{buy}}(t) + V_{\mathrm{sell}}(t)},
\]
ranging from 0 (one-sided pressure) to 1 (perfectly balanced flow). For informed-flow detection, low two-sidedness combined with high $\mathrm{VR}(6)$ is the prototypical signature: persistent, directionally consensual pressure that does not reverse---consistent with a small set of informed agents pushing the price toward the eventual outcome.

\medskip

These two diagnostics are taken from the Signal Credibility Index of \citet{nechepurenko2026focal}, where they are used to assess whether a price move will acquire \emph{social} authority. Here we use them, alongside on-chain wallet features, for the parallel question of whether a price move reflects \emph{informational} provenance. The shared underlying intuition is that durable, directional, broadly-sourced repricing is qualitatively different from transient, contested, narrowly-sourced repricing; the two frameworks differ in which downstream consequence they map this distinction onto.

\subsection{Trade-size and time-clustering features}
\label{sec:micro-extra}

In addition to the three families above, we extract two empirical features that proved useful in pilot examination of public insider cases. Both are simple to compute and contribute distinct information at the margin.

\paragraph{Trade-size kurtosis.} The empirical kurtosis of trade sizes in a rolling window. Informed traders frequently break a target position into a small number of moderately sized trades, producing higher kurtosis than the retail baseline.

\paragraph{Hawkes-style self-excitation.} A simple intensity measure of clustering in trade arrivals. Informed flow tends to be self-exciting: a single informed trade is often followed by additional same-direction trades from the same or related wallets within minutes. We fit a one-dimensional univariate Hawkes process \citep{hawkes1971spectra} to the trade arrival times in the rolling window and report the estimated branching ratio.

\subsection{Wallet features at inference time}
\label{sec:micro-wallets}

The wallet-level features defined in \Cref{sec:wn} are causal and can be computed at inference time. For each new trade entering the rolling window, we look up the trading wallet's age, prior-market participation count, funding-source concentration (cached from prior queries), and entry timing relative to the market's resolution date. These per-trade values are aggregated within the window to form market-level wallet features.

\subsection{Feature vector and detector}
\label{sec:micro-vector}

For each active market $M$ at inference time $t$, we form a feature vector
\[
x(M, t) \;=\; \bigl(\mathrm{OI}_{5\text{m}}, \mathrm{OI}_{15\text{m}}, \mathrm{OI}_{1\text{h}}, \VPIN, \hat{\lambda}, \mathrm{VR}(6), \mathrm{TS}, \kappa_{\text{size}}, \beta_{\text{Hawkes}}, \overline{\WN}, c_{\text{cat}}, c_{\text{liq}}, c_{\text{TTR}}\bigr),
\]
where $c_{\text{cat}}$ encodes the market category, $c_{\text{liq}}$ encodes the liquidity tier, and $c_{\text{TTR}}$ encodes the time to resolution. The detector outputs a calibrated probability
\[
\hat{q}(M, t) \;=\; \Prob\!\bigl[y_{\mathrm{bin}}(M) = 1 \,\big|\, x(M, t)\bigr]
\]
that, conditional on the current observable state, the market will eventually be classified as informed-flow-positive under the labeling rule of \Cref{sec:ils-detector}. The class of detector model is intentionally kept simple: gradient-boosted trees on the engineered feature vector, with isotonic-regression calibration applied to the model output \citep{platt1999probabilistic, niculescu2005predicting}. We see no \emph{a priori} reason to expect that a sequence model would outperform this baseline given the relatively small training-set size, but this is an empirical question that we will address in the companion section.

\subsection{Statistical power and ground-truth scarcity}
\label{sec:micro-power}

A binding practical constraint on the empirical validation is the small number of unambiguously-labeled insider cases available in the public record. Combining documented episodes from \citet{mitts2026iran}, public reporting, and direct on-chain inspection, we estimate roughly 10--30 cases of high-confidence informed flow distributed across our three target categories during the two-year sample window. This is a small training set by the standards of contemporary supervised learning.

We follow the power-analysis methodology of \citet{nechepurenko2026foresight} to assess what magnitude of detector skill is reliably distinguishable at this sample size. For a binary classifier with true sensitivity $\pi_1$ on positives and false-positive rate $\pi_0$ on negatives, distinguishing the detector from a no-skill baseline at significance $\kappa = 0.05$ and power $0.80$ requires roughly
\[
n_{\text{pos}} \;\gtrsim\; \frac{(z_{1-\kappa} + z_{0.80})^2 \cdot \pi_1 (1 - \pi_1)}{(\pi_1 - \pi_0)^2}.
\]
For $\pi_1 = 0.7, \pi_0 = 0.2$, this yields $n_{\text{pos}} \approx 14$---comfortably within reach for a coarse skill claim. For finer claims, e.g. distinguishing $\pi_1 = 0.6$ from $\pi_1 = 0.5$, the required sample grows by an order of magnitude.

The implication is threefold. First, the power calculation assumes a labelled positive set is available; the pilot of \Cref{sec:pilot} establishes that the present version of the score does not produce such a set on the FFIC inventory under either the resolution-anchored proxy or the article-derived anchor for the markets in scope, because the inventory's documented cases are predominantly deadline-resolved. The deadline-ILS extension of \Cref{sec:extension} is therefore a precondition for the power analysis to apply at the FFIC inventory's scale. Second, conditional on that extension, the present work would claim to detect strong leakage signatures (e.g., the Iran or Year-in-Search episodes) with the expected high power; finer claims about marginal cases must wait for either more public episodes or for the productionized monitoring system itself to surface candidate cases for retrospective labeling. Third, the proper-scoring-rule framework of \Cref{sec:ils-murphy} provides a continuous label that does not require binary insider/non-insider classification, materially relaxing the labeled-data constraint at the cost of higher label noise.

\subsection{Implementation status}
\label{sec:micro-impl-status}

The system described above is implemented and partially operational. \Cref{tab:impl-status} summarizes the state of the data and analytical layers as of this writing. The historical-backfill stage is complete; the analytical layer ($\Tnews$ recovery, ILS computation, microstructure features) is in active development; the real-time stage and detector training are sequenced for the immediately following work. The numbers in \Cref{tab:impl-status} should be read as a snapshot of an actively evolving system, not as final empirical results.

\begin{table}[t]
\centering
\caption{ForesightFlow data and analytical pipeline status. Volume thresholds reflect the empirical CLOB-coverage cutoff documented in \Cref{sec:data-sample}: below approximately \$50K, the Polymarket subgraph reliably returns no trade history because the corresponding markets resolve through automated oracles without on-book trading. The dash entries are stages whose execution is in progress or sequenced for the next development cycle.}
\label{tab:impl-status}
\small
\renewcommand{\arraystretch}{1.25}
\begin{tabularx}{\linewidth}{@{}p{4.6cm}Xc@{}}
\toprule
\textbf{Component} & \textbf{Coverage / output} & \textbf{Status} \\
\midrule
Market metadata (Gamma)            & 911{,}237 markets, 2020--2026, 76 monthly buckets   & \checkmark complete \\
Resolution outcomes (Gamma)        & 865{,}725 resolved with binary outcome              & \checkmark complete \\
Category classification            & Keyword-based; 127K resolved in target categories   & \checkmark complete \\
Resolution-type classification     & 11{,}200 markets typed; 1{,}145 event-resolved      & \checkmark complete \\
Trade history (subgraph)           & 17.9M trades across 10{,}410 markets ($\geq$\$50K)  & \checkmark complete \\
Wallet inventory (subgraph + chain)& 796K unique wallet addresses, on-chain seeded       & \checkmark complete \\
Trade-VWAP price reconstruction    & Implemented; 725 markets price-series available     & \checkmark complete \\
$\Tnews$ recovery (proxy)          & Resolution-anchored proxy on event-resolved          & \checkmark pilot    \\
ILS computation under proxy        & Pilot $n=725$ + control $n=683$, no separation       & \checkmark pilot    \\
Resolution evidence URLs (UMA)     & 0\% on event-resolved (admin-resolved by design)    & --- (deferred)      \\
$\Tnews$ recovery (Tier 1: article)& Blocked by UMA URL absence on event-resolved        & --- (blocked)       \\
$\Tnews$ recovery (Tier 2: GDELT)  & Sequenced --- batch recovery on FFIC validation     & --- (priority)      \\
$\Tnews$ recovery (Tier 3: LLM)    & Proof-of-concept complete on Barak; ILS robust       & \checkmark POC      \\
FFIC eligibility audit             & 0 of 24 markets eligible under original scope        & \checkmark complete \\
Deadline-ILS specification         & Section 7 specification, paper-as-spec               & \checkmark complete \\
Deadline-ILS implementation        & Code path \texttt{compute\_ils\_deadline}, scope check & \checkmark complete \\
Hazard-rate estimation             & Per-category MLE; 2 of 3 fits adequate               & \checkmark partial  \\
FFIC trade-history backfill        & 20 of 24 markets recovered; 4 election markets blocked & \checkmark partial  \\
$\Tevent$ recovery (Tier 3 LLM)    & Generalized to deadline-YES; FFIC batch complete     & \checkmark complete \\
Deadline-ILS empirical evaluation  & Iran-Apr30 case study; 5 refinements surfaced        & \checkmark partial  \\
Microstructure features            & Code paths defined; computation deferred to Task 04 & ---                  \\
Real-time streaming                & Architecture defined; deferred to Task 05           & ---                  \\
Detector training                  & Sequenced after labelled set is materialized        & ---                  \\
\bottomrule
\end{tabularx}
\end{table}

A practical reason to publish the pipeline status alongside the methodology is that the bottleneck for empirical validation is, at this stage, neither algorithmic nor methodological but operational: the rate at which $\Tnews$ values can be reliably recovered for resolved markets in the target categories. The framework's empirical claims will be commensurate with that recovery rate, which we will report explicitly when the analytical stage completes.

\subsection{End-to-end pipeline}
\label{sec:micro-pipeline}

\begin{figure}[t]
\centering
\resizebox{\linewidth}{!}{%
\begin{tikzpicture}[
  every node/.style={font=\scriptsize, align=center},
  src/.style={rectangle, draw, fill=blue!8, rounded corners=1.5pt, minimum width=2.4cm, minimum height=0.85cm, inner sep=2pt},
  proc/.style={rectangle, draw, fill=orange!18, rounded corners=1.5pt, minimum width=2.6cm, minimum height=0.85cm, inner sep=2pt},
  store/.style={cylinder, draw, shape border rotate=90, aspect=0.25, fill=green!12, minimum width=2.4cm, minimum height=1.0cm},
  output/.style={rectangle, draw, fill=red!10, rounded corners=1.5pt, minimum width=2.4cm, minimum height=0.85cm, inner sep=2pt},
  arrow/.style={-Latex, thick, gray!60!black},
  node distance=0.5cm and 0.7cm
]

\node[src] (gamma) {Polymarket\\Gamma API};
\node[src, right=of gamma] (clob) {Polymarket\\CLOB};
\node[src, right=of clob] (sg) {Polymarket\\subgraph};
\node[src, right=of sg] (uma) {UMA\\Oracle};
\node[src, right=of uma] (gdelt) {GDELT 2.0};
\node[src, right=of gdelt] (poly) {Polygonscan};

\node[proc, below=1.0cm of clob] (back) {Historical\\backfill};
\node[proc, below=1.0cm of uma] (stream) {Real-time\\stream};

\node[store, below=0.9cm of back] (db) {Postgres +\\TimescaleDB};
\node[store, below=0.9cm of stream] (cache) {Redis\\state cache};

\node[proc, below=0.9cm of db, fill=orange!28] (anal) {Analytics\\engine\\(\S\ref{sec:ils})};
\node[proc, below=0.9cm of cache, fill=orange!28] (det) {Detection\\engine\\(\S\ref{sec:micro})};

\node[proc, below=0.9cm of $(anal)!0.5!(det)$, fill=yellow!22, minimum width=4cm] (api) {API layer\\REST + WebSocket};

\node[output, below left=0.7cm and -0.2cm of api] (dash) {Dashboard\\(React)};
\node[output, below=0.7cm of api] (bot) {Telegram\\alerting};
\node[output, below right=0.7cm and -0.2cm of api] (pub) {Public\\read-only API};

\foreach \s in {gamma, clob, sg, uma} \draw[arrow] (\s) -- (back);
\foreach \s in {clob, sg, uma, gdelt, poly} \draw[arrow] (\s) -- (stream);

\draw[arrow] (back) -- (db);
\draw[arrow] (stream) -- (cache);
\draw[arrow] (stream) -- (db); 

\draw[arrow] (db) -- (anal);
\draw[arrow] (cache) -- (det);
\draw[arrow] (db.east) to[bend left=18] (det.north);

\draw[arrow] (anal) -- (api);
\draw[arrow] (det) -- (api);

\draw[arrow] (api) -- (dash);
\draw[arrow] (api) -- (bot);
\draw[arrow] (api) -- (pub);

\end{tikzpicture}%
}
\caption{End-to-end ForesightFlow pipeline. Data sources (top) feed historical backfill and real-time streaming paths; the analytics engine computes ILS labels and category statistics from the historical store, while the detection engine applies the trained model to the live feature stream. Outputs are exposed through a unified API to downstream consumers.}
\label{fig:pipeline}
\end{figure}
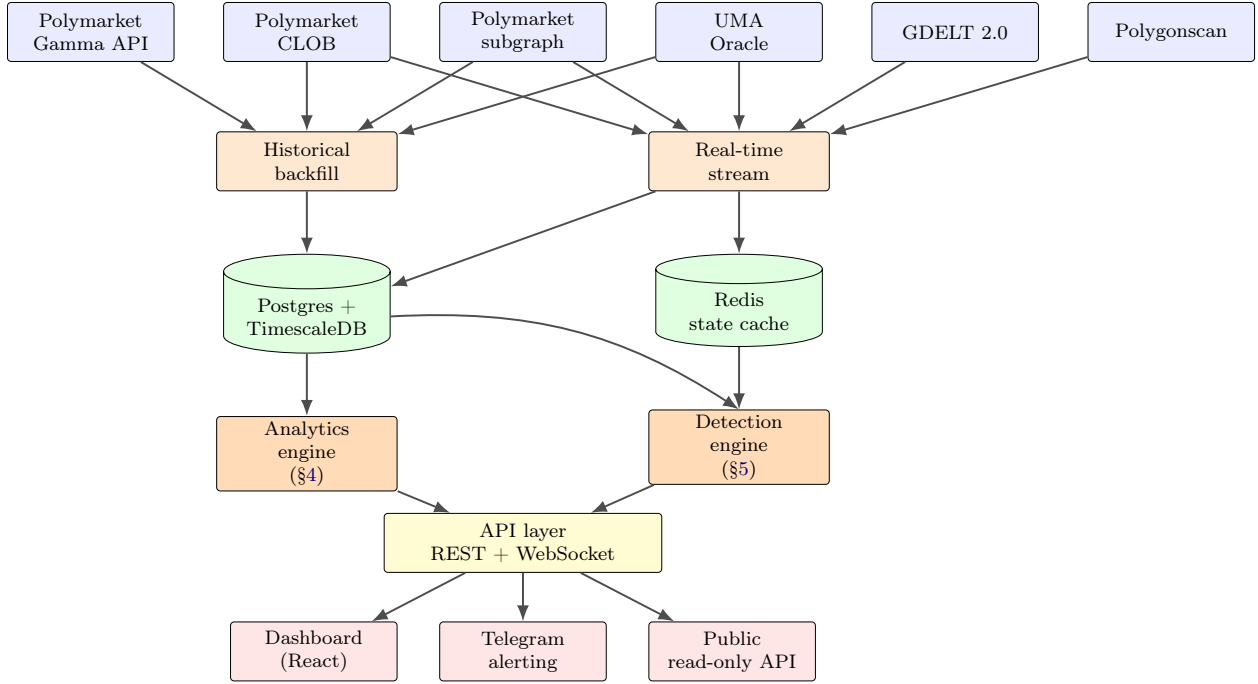

\Cref{fig:pipeline} summarizes the end-to-end flow from raw data sources to detector output and downstream consumers. The data layer is the union of Polymarket APIs, the UMA Optimistic Oracle, GDELT, and on-chain context from Polygonscan. Historical backfill populates the training database with computed ILS labels; the real-time stream populates per-market feature vectors at minute resolution. The analytics engine computes the labels and category-conditional priors from the historical store; the detection engine applies the trained model to the live feature stream. Detector output is exposed via a REST and WebSocket API to downstream consumers: a research dashboard, a Telegram alerting bot, and a public read-only API.

\section{Pilot Empirical Study and Methodological Lessons}
\label{sec:pilot}

This section reports a pilot empirical study run on $n = 725$ event-resolved markets in the three target categories. The study was originally framed as a first end-to-end validation of ILS as a label generator. The findings are negative: under the resolution-anchored proxy that the labelled set could feasibly be computed against, ILS does not separate the candidate informed-flow population from a matched control population. We nevertheless report the study in full, because the negative findings are the empirical basis for the scope conditions introduced in \Cref{sec:ils-scope} and for the next-stage methodological priorities. The honest summary of what we have learned is that ILS, as defined, is operationally interpretable only under conditions tighter than those originally proposed; and the absence of an article-derived $\Tnews$ is the binding constraint on present-stage validation.

\subsection{Setup}
\label{sec:pilot-setup}

The pilot scope is the 1{,}145 event-resolved markets identified in \Cref{sec:ils-typology} with volume $\geq \$50$K in the three target categories. Of these, 954 markets had trade history in the database after the Phase~3B subgraph backfill, and ILS was computable for 725 markets after applying the price-data and timestamp-validity filters. A minute-resolution price series was reconstructed from on-chain trades using a volume-weighted average within each minute (a trade-VWAP fallback to the Polymarket CLOB mid-price path described in \Cref{sec:data-polymarket}); CLOB mid-price was unavailable for the historical resolved-market sample because the live CLOB collector during Phase 3B was scoped to active markets only.

For the news timestamp $\Tnews$, in the absence of recoverable article-derived timestamps for admin-resolved markets (\Cref{sec:data-news}), we adopt the resolution-anchored proxy $\Tnews \equiv \Tres - 24\,\text{h}$. We treat this proxy as a working hypothesis to be evaluated, not as a substitute for a real news timestamp; one of the principal findings below is that the proxy is structurally misaligned with the events it is meant to anchor.

\subsection{Pilot ILS distribution and control comparison}
\label{sec:pilot-distribution}

\Cref{tab:pilot-vs-control} reports the central comparison. The pilot sample of 725 event-resolved markets is matched against a control sample of 683 unclassifiable markets (sports outcomes, behavioural-prediction markets, count markets) drawn from the same database with identical volume and trade-coverage filters. Both samples are scored under the same resolution-anchored proxy.

\begin{table}[t]
\centering
\caption{Pilot event-resolved sample versus matched unclassifiable control under the resolution-anchored $T_{\text{news}}$ proxy. The control population shows a higher positive-ILS rate than the pilot, a reversal of the prior expectation. Mann-Whitney $U$ test on the full ILS distributions yields $p = 1 \times 10^{-6}$; the $95\%$ bootstrap confidence interval on the difference of medians (control minus pilot) is $[+0.023, +0.066]$, entirely positive. The pilot does not separate from the control under this proxy.}
\label{tab:pilot-vs-control}
\small
\renewcommand{\arraystretch}{1.25}
\begin{tabularx}{\linewidth}{@{}Xrrr@{}}
\toprule
\textbf{Group}                            & \textbf{$n$} & \textbf{Median ILS} & \textbf{Positive ILS \%} \\
\midrule
Pilot (event-resolved target categories)  & 725          & $-0.084$            & 15.2\% \\
Control (unclassifiable, matched filters) & 683          & $-0.043$            & 21.4\% \\
\bottomrule
\end{tabularx}
\end{table}

The control sample, drawn from the population we expected \emph{a priori} to host less informed-flow activity, shows the higher positive-ILS rate. This is not consistent with the framework's prediction that event-resolved political and regulatory markets host elevated leakage relative to sports and behavioural markets. The two-sample test rejects the null of equal distributions, but in the wrong direction.

The mechanism becomes legible once the resolution structure of each population is examined. For unclassifiable markets in this sample---predominantly sports outcomes and behavioural-count markets---the resolution event \emph{is} the news event. A football match ending, a tweet count completing, a Counter-Strike series concluding: in each case, the underlying observable that determines the market's resolution occurs at or very near $\Tres$, and the resolution-anchored proxy $\Tres - 24\,\text{h}$ legitimately captures a pre-news interval. For event-resolved political and regulatory markets, by contrast, the actual news event (an election called by major networks, a regulatory decision announced, a verdict rendered) typically precedes $\Tres$ by hours to days, and the resolution-anchored proxy captures a window that is partially or entirely \emph{post}-news. Pre-news leakage is invisible in this window because the leakage has already become public.

The headline finding of the pilot is therefore not that informed flow is absent from event-resolved markets, but that the proxy used to detect it is structurally misaligned with the event timing for this population. This is a negative result in the strict sense: the data, conditional on the proxy, do not support the prior expectation. It is also a methodological result in the constructive sense: it identifies the binding constraint that the next stage of empirical work must address.

\subsection{Proxy-sensitivity analysis}
\label{sec:pilot-proxy-sensitivity}

To assess whether tighter anchoring recovers a positive signal that the 24-hour proxy obscures, we recompute ILS on the pilot sample under three additional resolution-anchored offsets: $\Tres - 6\,\text{h}$, $\Tres - 2\,\text{h}$, and $\Tres - 1\,\text{h}$. \Cref{tab:proxy-sensitivity} reports the resulting distributions.

\begin{table}[t]
\centering
\caption{ILS distributions on the pilot sample under four resolution-anchored proxies. Tightening the proxy collapses the positive-ILS share to zero and inflates the share of $|\ILS| > 1$ extreme values. The Spearman rank correlation $\rho$ between the 24-hour and 1-hour proxies, computed on the 221 markets for which both are defined, is 0.542---moderate, not robust. The pilot's 15.2\% positive rate at the 24-hour offset does not survive tightening.}
\label{tab:proxy-sensitivity}
\small
\renewcommand{\arraystretch}{1.25}
\begin{tabularx}{\linewidth}{@{}lXrrrrr@{}}
\toprule
\textbf{Proxy} & & \textbf{$n$} & \textbf{Median} & \textbf{Pos.\ \%} & \textbf{$|\ILS| > 1$ \%} & $\rho_{24\text{h}}$ \\
\midrule
$\Tres - 24\,\text{h}$ & & 725 & $-0.084$ & 15.2\% & 13.9\% & --- \\
$\Tres - 6\,\text{h}$  & & 592 & $-0.134$ & 11.0\% & 19.3\% & 0.763 \\
$\Tres - 2\,\text{h}$  & & 316 & $-0.332$ & 0.0\%  & 25.3\% & 0.542 \\
$\Tres - 1\,\text{h}$  & & 221 & $-0.350$ & 0.0\%  & 27.1\% & 0.542 \\
\bottomrule
\end{tabularx}
\end{table}

The expected behaviour under genuine pre-news leakage is that ILS values rise as the anchor tightens toward the actual news event: a high-leakage market should concentrate its discriminative resolution in a progressively shorter pre-news window. The observed behaviour is the opposite. Positive ILS shares decline monotonically and reach zero at the two-hour and one-hour offsets, while extreme negative values proliferate. The Spearman rank correlation between the 24-hour and one-hour proxies is 0.542 on the 221 markets common to both, indicating that more than half of the rank variance is explained by anchor choice itself.

This is the empirical content of the anchor-sensitivity scope condition introduced in \Cref{sec:ils-scope}. ILS values that change qualitatively with offset are not robust signals; the pilot's apparent 15.2\% positive rate is not preserved under offset variation and does not pass the robustness check that the methodology now requires.

\subsection{Top-signal cluster: Epstein files disclosure markets}
\label{sec:pilot-epstein}

Three markets concerning whether named individuals would appear in the December 2025 Epstein files release scored ILS values of $0.93$, $0.64$, and $0.55$ under the 24-hour proxy. These were the highest values in the pilot's 24-hour distribution, and we reported them in earlier drafts of this work as candidate informed-flow signatures. A deep dive on the cluster reveals that two of the three are explained by the edge-effect regime of \Cref{sec:ils-scope}, and the third, while not edge-dominated, does not survive the proxy-sensitivity check. We report the analysis in full because it is, in our view, the cleanest concrete instance of the scope conditions in operation.

\paragraph{AOC and Sanders markets: edge-effect regime.} The market on whether Alexandria Ocasio-Cortez would be named in the release opened with $p_{\Topen} = 0.940$, drifted to $p(\Tnews) = 0.996$, and resolved YES; the Sanders market opened at $p_{\Topen} = 0.910$ and reached $0.968$ before $\Tnews$. In both cases the absolute pre-news price move is small (on the order of five percentage points), but the denominator $\Delta_{\text{total}} = p_{\Tres} - p_{\Topen}$ is also small (six and nine percentage points respectively), and the ratio is amplified into a near-unit ILS. Both markets violate the scope condition $|p_{\Topen} - 0.5| \leq 0.4$ introduced in \Cref{sec:ils-scope}. The high ILS values reflect the geometry of the formula in the high-consensus regime, not informational asymmetry.

\paragraph{Barak market: anchor sensitivity.} The market on whether Ehud Barak would be named opened at $p_{\Topen} = 0.170$, exhibited a substantial price-discovery period (a sell-off to $0.216$ on December 20 followed by a recovery to $0.529$ on December 21, both on elevated volume), and reached $p(\Tnews) = 0.629$ before resolving YES. The opening price satisfies the edge-effect scope condition, the absolute price move is large, and the trajectory has features consistent with informational arrival. However, the ILS value at the 24-hour proxy is $0.553$, and at the 6-hour proxy it is $-4.241$; the qualitative sign and magnitude are not preserved across anchor choices. The Barak market is consequently the highest-priority target for article-derived $\Tnews$ recovery (\Cref{sec:pilot-status}), but the proxy-based ILS values reported above cannot be interpreted as evidence of informed flow without that recovery.

\paragraph{Wallet-level inspection.} The three Epstein markets share a common dominant participant: a single wallet (\texttt{0x4bfb41d5\dots}) accounts for 80\% of pre-news YES volume in the AOC market, 18\% in the Sanders market, and 75\% in the Barak market, totalling \$34{,}034 in combined exposure. This wallet has traded over 5{,}100 distinct Polymarket markets since December 2022 and is consistent in profile with a professional liquidity-providing participant rather than a newly-created insider account. The AOC and Sanders positions, taken at average prices above $0.98$, have minimal directional profit potential; the Barak position, taken at an average price near $0.46$, has substantial directional profit potential and is the only one of the three that is economically consistent with informed positioning. Wallet-level features alone do not resolve the question for the Barak market; that resolution requires the article-derived $\Tnews$.

\subsection{Article-derived $T_{\text{news}}$: the Barak market as proof-of-concept}
\label{sec:pilot-barak-tier3}

The Barak market is the one cluster element that is not edge-effect-dominated and that exhibits a non-trivial price-discovery trajectory. It is therefore the natural target for a single-market proof-of-concept of article-derived $\Tnews$ recovery: if the proxy-based ILS value were itself the source of the apparent signal, replacing the proxy with the actual public-release timestamp should change the ILS value materially. We performed this substitution; it does not.

\paragraph{Recovered timestamp.} Cross-referencing five contemporaneous news sources (CNN, Al Jazeera, Times of Israel, NBC, ABC) with congressional press releases identifies the relevant public event as the December 18, 2025 release of 68 Epstein-estate photographs by the U.S.\ House Oversight Committee Democrats, one day before the Department of Justice's statutory deadline. A Barak photograph is in this batch. We therefore set $\Tnews = $ 2025-12-18T18:00 UTC, with confidence 0.90 (multi-source date agreement; intra-day time uncertainty $\pm 4$ hours). The proxy timestamp $\Tres - 24\,\text{h}$ falls on December 22 at 12:08 UTC, four days and eighteen hours \emph{later} than the actual release.

\paragraph{ILS comparison.} \Cref{tab:barak-ils-comparison} reports the side-by-side computation. The proxy-derived and article-derived ILS values are within two percentage points of each other, on opposite sides of zero from neither qualitative interpretation.

\begin{table}[t]
\centering
\caption{ILS for the Barak Epstein market under the proxy and article-derived $T_{\text{news}}$. The proxy is four days, eighteen hours later than the actual public release; the recovered ILS values nevertheless differ by less than two percentage points and yield the same qualitative interpretation. For this market, the proxy was not the binding error source.}
\label{tab:barak-ils-comparison}
\small
\renewcommand{\arraystretch}{1.25}
\begin{tabularx}{\linewidth}{@{}Xllr@{}}
\toprule
\textbf{Anchor} & \textbf{$T_{\text{news}}$} & \textbf{$p(T_{\text{news}})$} & \textbf{ILS} \\
\midrule
Resolution-anchored proxy        & 2025-12-22 12:08 UTC & 0.629 & 0.553 \\
Article-derived (LLM Tier 3)     & 2025-12-18 18:00 UTC & 0.643 & 0.570 \\
\midrule
$\Delta$                         & $-$4d 18h            & $+0.014$ & $+0.017$ \\
\bottomrule
\end{tabularx}
\end{table}

\paragraph{Wallet timing reclassification.} The substitution does, however, materially reclassify the wallet-level evidence. Of the fifteen wallets that appeared as ``early'' relative to the proxy $\Tnews$, eight first traded \emph{after} the actual December 18 release: their entries fall in the December 19--22 window, during which the market price oscillated between $0.22$ and $0.69$ as participants debated whether a photograph qualified as a ``newly released file'' under the resolution criteria. These wallets are reactive participants in a post-news price-discovery process, not informed pre-news positions. Only six wallets are genuinely pre-news under the article-derived anchor, and the dominant wallet (\texttt{0x4bfb41d5}, 5{,}115 markets traded since 2022) accounts for 92.6\% of pre-news YES volume; the remaining five pre-news wallets contribute approximately one thousand dollars combined.

\paragraph{Resolution-criteria arbitrage as a distinct phenomenon.} The December 20 price crash from $0.529$ to $0.216$ on 767 trades, followed by an immediate recovery to $0.529$ on December 21 with \$9{,}332 in volume, is consistent with a market debating a legal-interpretation question: does a photograph in which the named individual appears constitute material from the ``Epstein files'' for resolution purposes? This is a distinct phenomenon from informed trading on private information about the underlying event. We surface it because the trajectory of resolution-criteria arbitrage---a sharp, high-volume reversal driven by interpretation of contract language rather than by new information about the world---can be confused with informed flow at the level of the price series alone, and we expect it to recur on Polymarket in legally ambiguous market formulations.

\paragraph{What the proof-of-concept shows.} The substitution yields three findings, each of which alters our prior reading of the case. First, ILS is essentially unchanged at the proper anchor, which means proxy quality was not the primary cause of the moderate-positive value: ILS $\approx 0.57$ for the Barak market is a robust feature of the price trajectory under either anchor, not a proxy artefact. Second, the wallet-level evidence with the correct anchor narrows the set of plausibly-informed participants from fifteen to six and concentrates 92.6\% of pre-news volume in a single professional wallet whose Polymarket history is inconsistent with newly-created insider activity. Third, the most anomalous price feature in the trajectory, the December 20 crash, has a clean non-informational explanation: it occurs after the public release, not before it, and is consistent with on-chain disagreement about contract language. The combination of these three findings is that ILS $= 0.570$ on the Barak market with the correct anchor, taken together with the wallet inventory at that anchor, does not by itself constitute evidence of informed trading. It is consistent with a moderately-informative market reaching its eventual resolution through a combination of staged public information arrival and resolution-criteria arbitrage. We retain the Barak market in the validation inventory because the trajectory remains the cleanest one to study; we do not claim it as a positive identification of informed flow.

\subsection{Validation-set eligibility audit and the deadline-resolved structure of documented insider cases}
\label{sec:pilot-ffic-audit}

Before scaling the Barak proof-of-concept of \Cref{sec:pilot-barak-tier3} to a batch article-derived $\Tnews$ recovery across the full FFIC inventory, we audited the inventory against the scope conditions established earlier in this work. The audit checks five requirements jointly: presence of trade history in the database (at least 100 trades), classification as event-resolved by the resolution-typology classifier of \Cref{sec:ils-typology}, opening price in the interpretable range $p_{\Topen} \in [0.1, 0.9]$ per the edge-effect scope condition of \Cref{sec:ils-scope}, total information move $|\Delta_{\text{total}}| \geq 0.05$ per the trivial-resolution condition, and a non-null resolution timestamp. \Cref{tab:ffic-audit} reports the result.

\begin{table}[t]
\centering
\caption{FFIC inventory audit against the scope conditions of \Cref{sec:ils-scope,sec:ils-typology}. Of 24 markets across the eight documented cases, none simultaneously satisfy all conditions. The dominant exclusion is resolution-type: 21 of 24 markets are classified as deadline-resolved or unclassifiable rather than event-resolved. Sixteen of the 24 additionally have no trade history in the database due to subgraph-indexer limits at the highest volume tier or to gaps in the Phase 3B backfill window.}
\label{tab:ffic-audit}
\small
\renewcommand{\arraystretch}{1.25}
\begin{tabularx}{\linewidth}{@{}Xrr@{}}
\toprule
\textbf{Exclusion reason} & \textbf{$n$} & \textbf{\% of FFIC} \\
\midrule
Resolution-type not event-resolved (deadline / unclassifiable / null) & 21 & 87.5\% \\
No trade history in database ($n_{\text{trades}} < 100$)               & 16 & 66.7\% \\
No market\_label record (resolution-type unrecoverable)                & 19 & 79.2\% \\
Edge-effect violation ($p_{\Topen} > 0.9$)                              & 2  & 8.3\%  \\
\midrule
\textbf{Eligible markets (all conditions satisfied)}                   & \textbf{0} & \textbf{0\%} \\
\bottomrule
\end{tabularx}
\end{table}

The audit returns zero eligible markets. Reasons compound across the inventory: the four top-line 2024 U.S.\ presidential markets ($1.5$B and $1.0$B in volume respectively) lack trade history due to The Graph indexer-capacity limit documented in \Cref{sec:limitations}; the FTX, SBF, and Bitcoin ETF markets lack trade history due to gaps in the subgraph backfill window; the 2026 U.S.--Iran conflict and Maduro / Venezuela cluster markets are correctly classified as deadline-resolved by the typology classifier; the SBF-sentenced and Romanian-election markets fail the edge-effect condition with opening prices of $0.97$ and $0.99$ respectively.

The single dominant reason is the resolution-type mismatch. We treat this as the central finding of the audit and as a methodological observation in its own right.

\paragraph{Documented insider cases on Polymarket are systematically deadline-resolved.} Inspection of the public-reporting record that motivates each FFIC case reveals a consistent pattern. Documented insider activity on Polymarket has been reported on contracts of the form ``Will event $X$ occur by date $Y$?''---an Iran strike on Israel by November 8, U.S.\ entry into Iran by April 30, Maduro in U.S.\ custody by January 31, the U.S.\ invading Venezuela, the Bitcoin ETF approved by January 15. The reported trades are pre-event purchases of YES shares that pay out when the event occurs and the deadline is satisfied; the corresponding markets resolve YES because the event happened, or NO because the deadline expired without the event. In either case, the contract is structurally a deadline market.

This is not coincidence. The deadline structure is precisely what makes a market actionable for a participant with private information about the timing of an upcoming event: the contract has a fixed expiration, the YES leg is asymmetrically rewarded if the event happens on time, and the price during the pre-event window is a direct function of the market's aggregated belief about whether the event will occur. An insider with reliable information that the event is imminent has a clean position to enter; an insider with information that the event has been deferred or cancelled has the symmetric position on the NO leg. The cases reported by \citet{mitts2026iran} and the cases assembled in our own FFIC inventory all instantiate this pattern.

\paragraph{Implication for ILS scope.} The score as defined in \Cref{sec:ils-def} and refined in \Cref{sec:ils-typology,sec:ils-scope} is restricted to event-resolved markets, where $\Tnews$ is the timestamp of a public mention of an observable event. For deadline-resolved markets, the canonical news event is either the occurrence of the event (which collapses $\Tnews$ to a moment shortly before $\Tres$) or its non-occurrence at the deadline (where $\Tnews \equiv \Tres$ and ILS is identically zero by construction). Neither reading captures the informed-flow signature that the empirical literature has documented: pre-event positioning by participants who know an event is about to happen.

The audit therefore exposes a structural gap between the score as currently defined and the population of markets in which the phenomenon of interest is empirically attested. Closing this gap is not a question of better data or tighter proxy; it requires extending the score to deadline-resolved cases. We sketch such an extension in \Cref{sec:extension} and treat its empirical evaluation as the binding next step.

\subsection{Status and next steps}
\label{sec:pilot-status}

The pilot's findings, including the Barak proof-of-concept, motivate the following sequencing of subsequent empirical work.

\paragraph{Article-derived $T_{\text{news}}$ is necessary but not sufficient.} The Barak case demonstrates that proxy quality was not the binding constraint on validation, contrary to our prior expectation. Replacing the proxy with the article timestamp shifts ILS by less than two percentage points and does not invert the substantive reading of the case. This implies that the next-stage empirical work cannot rely on $\Tnews$ recovery alone to materialize the labelled set; ILS plus the wallet-level features, jointly, must be the decision basis. We retain Tier 2 (GDELT) and Tier 3 (LLM-assisted) recovery as priorities because the wallet-timing reclassification they enable is itself a first-order improvement over the proxy.

\paragraph{The scope conditions introduced in \Cref{sec:ils-scope} are mandatory, not optional.} Reporting raw ILS values on markets that violate the edge-effect or anchor-sensitivity conditions produces results that cannot be cleanly interpreted, as the AOC and Sanders cases illustrate. We adopt the rule that any future ILS-based claim is accompanied by the corresponding scope-condition check.

\paragraph{Resolution-criteria arbitrage warrants its own treatment.} The December 20 Barak crash is, on inspection, a distinct empirical phenomenon---a sharp, high-volume reversal driven by interpretation of contract language rather than by external information arrival. It is invisible to ILS as currently specified and could plausibly be flagged by a separate detector trained on within-market reversal signatures conditional on no concurrent news event. We mark this as a separable research question that the framework as currently specified does not address.

\paragraph{The control population is itself informative.} The unclassifiable population's higher positive-ILS rate under the 24-hour proxy is not a methodological failure of the framework as a whole; it identifies a class of markets (sports, behavioural) for which the resolution-anchored proxy is structurally valid and where short-window information dynamics are detectable. This is a separable research question, parallel to the framework's primary aim, and we mark it as future work.

\paragraph{The detector itself has not been trained.} The pilot reports label distributions, not detector performance. With the methodological constraints now made explicit, training proceeds in the next stage on a sample restricted by the scope conditions and accompanied by wallet-level features. The next-stage data input was originally planned to be a batched article-derived $\Tnews$ recovery across the FFIC validation set; the audit reported in \Cref{sec:pilot-ffic-audit} identified a structural obstacle to that plan, which we treat as a separable methodological finding rather than a delay, and which motivates the deadline-ILS extension presented in \Cref{sec:extension}.

\section{Extending ILS to Deadline-Resolved Markets}
\label{sec:extension}

The pilot of \Cref{sec:pilot} establishes the operational reading of ILS for event-resolved markets. The audit of \Cref{sec:pilot-ffic-audit} establishes that the documented Polymarket insider-trading cases of practical interest are predominantly deadline-resolved and therefore lie outside that scope. This section sketches an extension that closes the gap. We do not claim a complete formal treatment in this draft; the goal is to specify the extension precisely enough that the scope conditions, the choice of $\Tnews$, and the connection to the existing framework are unambiguous, and to describe what the empirical evaluation would look like once executed.

\subsection{The deadline market as a Bayesian belief tracker}
\label{sec:extension-bayes}

Consider a deadline market on the question ``Will event $E$ occur by deadline $D$?''. Let $\theta \in [0, 1]$ denote the participants' marginal belief at any given time that $E$ has occurred or will occur before $D$. Under a calibrated market, $p(t) \approx \theta_t$, the price tracks the aggregated belief.

In the absence of new information, $\theta_t$ is determined by the survival of the deadline window. A simple Bayesian baseline takes the form
\[
\theta_t \;=\; \theta_{\Topen} \cdot \frac{S(D - t \mid \Topen)}{S(D - \Topen \mid \Topen)},
\]
where $S(\tau \mid \Topen)$ is the survivor function of the time-to-event distribution conditional on the information available at $\Topen$. Without informative news, $\theta_t$ is monotonically non-increasing on $t \in [\Topen, D]$ until the moment the event occurs (if it does), at which point the price jumps to $1$.

For empirical implementation we adopt a constant-hazard parametric form, $S(\tau) = e^{-\lambda \tau}$, with the rate $\lambda$ fitted on a comparable sample of resolved deadline markets in the same target category that resolved YES. Under this specification, $\theta_t$ admits the closed form
\[
\theta_t \;=\; \theta_{\Topen} \cdot \frac{1 - e^{-\lambda(D - t)}}{1 - e^{-\lambda(D - \Topen)}}.
\]
The constant-hazard assumption is a deliberately weak baseline: it asserts only that, conditional on no information arrival, the conditional probability that the event occurs in the next instant given that it has not yet occurred is constant. We use it because (i) it requires a single empirically-fit parameter, (ii) it is the standard memoryless prior in the absence of further structure, and (iii) it does not assume any non-trivial information about event timing within the window. Tighter parametric forms---for instance, fitting $\lambda$ as a function of remaining time-to-deadline, or using Weibull rather than exponential survival---are natural refinements once the empirical evaluation surfaces deviations from the constant-hazard baseline.

Two qualitatively distinct events can move the price away from this passive trajectory. The first is occurrence of $E$ itself, observed publicly: the price jumps to $1$ at $\Tevent$. The second is informational arrival without occurrence: a public news event materially updates participants' belief about whether $E$ is more or less likely to occur in the remaining window. Both are publicly observable; neither corresponds to informed flow as we have defined it.

Informed flow in a deadline market takes a third form: a participant with private information about $\Tevent$ trades ahead of the public observation of the event. Concretely, the participant buys YES at price $p(t_\text{insider}) < 1$ during the window $[t_\text{insider}, \Tevent]$, where $t_\text{insider}$ is the moment private information becomes available. The participant's profit is realized at $\Tevent$ when the price jumps to $1$. This is precisely the trading pattern documented by \citet{mitts2026iran} for the October 2024 Iran-strike markets: pre-event purchases of YES shares at low prices that paid out when the strike occurred and the markets resolved YES.

\subsection{Definition: deadline-ILS}
\label{sec:extension-definition}

Let $\Tevent$ denote the timestamp of public observation of $E$, defined when $E$ occurs and undefined when the deadline expires without occurrence. For a deadline-resolved market in which $\Tevent$ is well-defined and falls within $[\Topen, D]$, define
\[
\Delta^{\text{dl}}_{\text{pre}} \;=\; p(\Tevent^-) - \theta_{\Topen}, \qquad
\Delta^{\text{dl}}_{\text{total}} \;=\; p_{\Tres} - \theta_{\Topen}, \qquad
\ILS^{\text{dl}}(M) \;=\; \frac{\Delta^{\text{dl}}_{\text{pre}}}{\Delta^{\text{dl}}_{\text{total}}}.
\]
Here $p(\Tevent^-)$ is the market price immediately before public observation of $E$, $\theta_{\Topen}$ is a baseline-belief reference (specified below), and $p_{\Tres} \in \{0, 1\}$ is the binary resolution outcome.

The structure is parallel to the event-resolved definition of \Cref{sec:ils-def}: the score asks how much of the move from baseline to resolution had occurred before the public-information arrival, which in this setting is the public observation of the event itself. The denominator is anchored to a baseline belief rather than to the raw opening price, in principle, because in a deadline market the opening price already reflects the prior on $\Tevent$ and is not in general a no-information benchmark.

\paragraph{Specifying the baseline $\theta_{\Topen}$.} The choice of baseline is the central design parameter of the extension. Three options are available: (i) the observed opening price, $\theta_{\Topen} \equiv p(\Topen)$; (ii) a parametric Bayesian prior derived from a survival-function model; (iii) an empirical base rate computed across a population of comparable deadline markets. We adopt option (i) for the present draft and treat (ii) and (iii) as future refinements. Two considerations motivate this choice. First, option (i) is conservative in the direction that matters for detection: it under-attributes pre-event ILS in cases where informed trading has already begun by the moment of market opening, producing false negatives rather than false positives. False positives are the more costly error type for an academic claim about insider trading, and we prefer the conservative bias. Second, for the high-volume political and regulatory deadline markets that dominate the FFIC inventory, opening prices are typically reasonable consensus priors rather than uniform $0.5$ baselines, which makes option (i) operationally close to a thoughtful empirical base rate without requiring its computation. We return to options (ii) and (iii) as future work in \Cref{sec:extension-empirical}.

\paragraph{Survival-function model and the time-to-event distribution.} Although the baseline $\theta_{\Topen}$ in the present draft does not require an explicit survival model, the underlying belief-tracker formulation of \Cref{sec:extension-bayes} does: the passive trajectory $\theta_t$ in the absence of news is determined by the survivor function $S(\tau \mid \Topen)$ of the time-to-event distribution. We adopt a constant-hazard (exponential) parametric form, $S(\tau) = \exp(-\lambda \tau)$, with the hazard parameter $\lambda$ estimated separately by category from the population of resolved deadline markets in our corpus that resolved YES with a recovered $\Tevent$. Per-category estimation is necessary because the empirical event rates differ substantially across our three target categories: military / geopolitical deadline markets concentrate in short lifetimes (hours to days) while regulatory and corporate-disclosure markets extend over weeks to months. A single global $\lambda$ would under-fit both populations. The empirical procedure is described in \Cref{sec:extension-empirical}.

\paragraph{Recovering $\Tevent$.} For YES-resolved deadline markets, $\Tevent$ is recovered through the same article-derived timestamp pipeline used for $\Tnews$ in event-resolved markets (\Cref{sec:data-news}, Tier 1--3). The retrieval target differs---``when did $E$ first publicly happen'' rather than ``when was $E$ first publicly mentioned''---but the operational pipeline is identical, and the same multi-source verification logic applies. We unify these two cases at the implementation level: a single \texttt{tier3} command branches on the resolution typology of \Cref{sec:ils-typology} to determine which timestamp it is recovering.

\paragraph{NO-resolved deadline markets.} For deadline markets in which the deadline expires without occurrence ($\Tevent$ undefined, $p_{\Tres} = 0$), the canonical $\Tnews$ for informed-flow detection is $\Tdead$, the deadline-expiration time, and ILS is identically zero by construction: the market resolves at the deadline because nothing happened, and there is no pre-deadline information arrival to anticipate. Informed flow on these markets, when present, takes the form of pre-deadline NO purchases by participants who knew the event would not occur; the corresponding diagnostic is the symmetric definition with the YES leg replaced by NO. Empirically, the case of an insider with information that an event will \emph{not} occur is rarer in the FFIC record than the symmetric YES case, but it is not absent (e.g., regulatory decisions held back through a planned deadline), and the symmetric formulation makes the detection invariant to this asymmetry.

\subsection{Scope conditions and connection to the existing framework}
\label{sec:extension-scope}

The deadline-ILS inherits the scope conditions of \Cref{sec:ils-scope} unchanged. The edge-effect condition is applied to $\theta_{\Topen}$ as defined above; under the choice $\theta_{\Topen} \equiv p(\Topen)$ adopted in this draft, this is equivalent to the original condition $|p(\Topen) - 0.5| \leq 0.4$ on the opening price, which we re-impose without modification. The trivial-resolution condition and the anchor-sensitivity condition are unchanged. The resolution-typology classifier of \Cref{sec:ils-typology} is augmented: deadline markets are now \emph{partially} in scope, specifically those resolving YES with a recoverable $\Tevent$ and those resolving NO with informed NO-side flow detectable via the symmetric definition.

The Murphy-decomposition reading of \Cref{sec:ils-murphy} extends naturally: $\ILS^{\text{dl}}(M)$ is approximately the share of the market's resolution component that was accumulated before public observation of $E$, where resolution is now measured against the baseline forecast at $\Topen$ rather than the marginal outcome rate.

\subsection{Empirical programme}
\label{sec:extension-empirical}

The deadline-ILS extension makes the FFIC inventory addressable. The October 2024 Iran-strike markets, the 2026 U.S.--Iran conflict cluster, the Maduro / Venezuela cluster, and the Bitcoin ETF approval markets are all deadline contracts with recoverable $\Tevent$ for the YES-resolved cases or recoverable $\Tdead$ for the NO-resolved cases. Implementation requires three coordinated steps. First, the Phase 3B subgraph backfill must be extended to recover trade history for the markets currently missing from the database, which is a separable infrastructure task. Second, the article-derived timestamp recovery already validated on the Barak market for $\Tnews$ (\Cref{sec:pilot-barak-tier3}) must be generalized to recover $\Tevent$ for deadline-YES markets and to verify $\Tdead$ for deadline-NO markets; the same Tier 3 pipeline supports all three with a market-type dispatch. Third, the constant-hazard rate $\lambda$ must be fitted on the comparable population for each target category. All three components have been implemented; the empirical evaluation, including the per-category hazard fits, the single-case ILS$^{\text{dl}}$ computation on the 2026 U.S.--Iran conflict cluster, and the methodological refinements surfaced by the analysis, is the subject of the companion paper \citep{nechepurenko2026foresightflow_empirical}.


\section{Summary and Next Steps}
\label{sec:summary}

This paper has set out the methodological foundation of the ForesightFlow framework for quantifying informed flow in active prediction markets. Five contributions structure the work. First, the task is reformulated from post-hoc forensics to real-time inference, with the architectural consequence that all features used at inference time must be causal in the available information. Second, labels are generated via the Information Leakage Score, computed on resolved markets in three pre-specified categories, with an explicit Murphy-decomposition reading that connects ILS to the proper-scoring-rule literature (\Cref{sec:ils-murphy}). Third, the detector combines classical microstructure measures, adapted to the bounded discrete setting of binary outcome markets, with on-chain wallet features unique to decentralized platforms; two of the microstructure diagnostics (variance ratio and two-sidedness) are drawn directly from the Signal Credibility Index of \citet{nechepurenko2026focal}, illustrating that a shared microstructural toolkit can serve both the social-authority and the informational-provenance questions. Fourth, the system is implemented as a production pipeline with a unified data layer, dual backfill and streaming paths, and downstream consumers spanning research, alerting, and public API. Fifth, the original ILS, restricted to event-resolved markets, is extended to a deadline-ILS variant that addresses the population in which insider trading has been empirically documented (\Cref{sec:extension}).

\paragraph{Companion empirical paper.} The companion paper \citep{nechepurenko2026foresightflow_empirical} carries the deadline-ILS extension through to an end-to-end empirical evaluation on the 2026 U.S.--Iran conflict cluster of the FFIC inventory, including per-category hazard-rate fits, a single-case ILS$^{\text{dl}}$ computation, cross-market wallet analysis, and the methodological refinements that the evaluation surfaces. We refer the reader to that paper for any concrete numerical claim about deadline-resolved markets and treat its findings as the primary empirical evidence accompanying the methodology developed here.

\paragraph{Next-stage research programme.} Three lines of follow-on work flow from the methodology and from the companion paper's empirical findings jointly. The first is infrastructure: continuous CLOB price collection and continuous per-trade collection from $\Topen$ for all markets in the target categories, both required to convert wallet-level features from a settlement-arbitrage diagnostic into a coordination-signal diagnostic. The second is methodology: sub-categorization of the regulatory-decision target category to remove the bimodal hazard-rate distribution observed in the companion paper, and an expanded corporate-disclosure hazard-rate sample. The third is detector training itself, which has not been undertaken in either paper of the present series: with the labelled set materialized via the deadline-ILS extension and the wallet inventory reconstructed via continuous trade collection, the detector architecture of \Cref{sec:micro-vector} can be trained on the joint feature space and evaluated against the operating points relevant to surveillance applications. Following the cumulative-evaluation methodology of \citet{nechepurenko2026foresight}, detector performance accumulates statistical power as the system runs in production rather than from a one-shot retrospective benchmark; the small number of unambiguously labelled cases requires this longitudinal approach.

\subsection{Limitations and threats to validity}
\label{sec:limitations}

We close the methodology presentation by stating, explicitly, the limitations that constrain the claims of this paper.

\paragraph{Regularity conditions on the Murphy interpretation.} The bridge from ILS to the resolution component of the Brier score (\Cref{rem:ils-as-resolution}) holds only under regularity conditions that are non-trivial in practice. Two conditions deserve emphasis. First, pre-news price moves must be unbiased estimators of the conditional outcome probability; if the market is systematically miscalibrated (for example, exhibiting a favourite-longshot bias), the proportionality between $\Delta_{\mathrm{pre}}$ and $\mathrm{RES}(\Tnews)$ holds only approximately and with a category-specific scale factor. Second, the binning argument underlying the Murphy decomposition assumes stationarity within bins; for markets in which price drifts continuously, this is a coarse approximation. Empirically, the consequence is that ILS values near zero or near one are reliable as ordinal indicators, but intermediate values should be interpreted as a rough fraction rather than a precise proportion of front-loaded resolution. The companion empirical paper \citep{nechepurenko2026foresightflow_empirical} examines this approximation error in concrete cases.

\paragraph{Label noise from $\Tnews$ recovery.} The hierarchical $\Tnews$ recovery procedure of \Cref{sec:data-news} produces labels of varying quality. UMA proposer evidence URLs are authoritative when available but are not always present; GDELT keyword matching introduces both false positives (irrelevant articles matching common keywords) and false negatives (the true first mention not in GDELT's source list). For the labelled validation set, each $\Tnews$ value is reviewed manually; for the larger backfill, ILS distributions are reported both with and without a confidence filter on the recovered timestamp.

\paragraph{Coverage of insider cases.} The FFIC inventory introduced in \Cref{sec:data-validation} addresses the small-N concern raised in earlier drafts: validation is anchored to eight documented episodes mapped to concrete Polymarket market identifiers. Two structural caveats apply. First, the inventory is biased toward episodes that reached enough public visibility to be reported, which almost certainly under-represents corporate insider trading. Second, the inventory is constructed retrospectively from cases where suspicion has already been raised; we do not claim that any trained detector built on this set will generalize to insider patterns qualitatively absent from it. The system is designed so that detection thresholds can be re-tuned as new public episodes accumulate, and the FFIC inventory is versioned to allow such updates to be tracked.

\paragraph{News-timestamp proxy quality and its limits.} The pilot in \Cref{sec:pilot} establishes empirically that the resolution-anchored proxy $\Tres - 24\,\text{h}$ is structurally misaligned with event-resolved political and regulatory markets at the population level. The Barak proof-of-concept in \Cref{sec:pilot-barak-tier3}, however, refines this picture: for an individual market that satisfies the edge-effect scope condition, replacing the proxy with the article-derived timestamp shifts ILS by less than two percentage points. We therefore identify two related but distinct constraints. The first is the population-level proxy structural mismatch, addressed by Tier 2 (GDELT) and Tier 3 (LLM-assisted) recovery. The second is that ILS at the proper anchor is itself not sufficient to identify informed flow: the Barak case yields ILS $\approx 0.57$ at the article-derived anchor and is nonetheless not interpretable as informed trading on its own. Future ILS-based claims combine $\Tnews$ recovery, scope-condition checks, and wallet-level feature evidence rather than relying on the score alone.

\paragraph{Validation-set / scope mismatch.} The audit of \Cref{sec:pilot-ffic-audit} established that the FFIC inventory is dominated by deadline-resolved markets that fall outside the scope of the original ILS. The deadline-ILS extension of \Cref{sec:extension} closes this gap at the level of methodology. The empirical evaluation of the extension---including the constraints under which it can be applied at scale---is reported in the companion paper \citep{nechepurenko2026foresightflow_empirical}.

\paragraph{Infrastructure constraints at extreme volume.} For the very largest markets in the FFIC inventory---the 2024 U.S.\ presidential top-line markets at \$1B+ each, the largest 2026 U.S.--Iran cluster contracts at hundreds of millions---the public Polymarket subgraph indexer on The Graph network returns errors indicating that no indexer has fully synchronized these markets, due to their size and event count. This is a hard infrastructure limit. Recovery would require either a self-hosted subgraph instance or direct event-log decoding via Polygon JSON-RPC, both of which add substantial operational complexity and are deferred to future work. As a consequence, the most consequential markets by raw volume are excluded from microstructure-level analysis, and the validation set effectively spans the \$100K to \$300M volume range with the \$1B+ tier left to future infrastructure investment.

\paragraph{Single-platform scope.} ForesightFlow as specified targets Polymarket only. Cross-platform analysis---comparing leakage signatures on Polymarket against those on Kalshi for the same underlying events---is a natural extension that is beyond the scope of the present work and that will require non-trivial harmonization across platforms with different microstructure and different regulatory regimes \citep{ng2026discovery, clinton2025prediction}.

\paragraph{Research programme.} ForesightFlow is the third component of an integrated research programme on the integrity, credibility, and informational structure of decentralized prediction markets. \citet{nechepurenko2026focal} addresses when a price signal carries enough social weight to coordinate elite and media behaviour. \citet{nechepurenko2026foresight} addresses how to evaluate AI forecasters on these markets without relying on PnL or trusted intermediaries. The present work addresses the third question in this triangle: when, on a given active market, is the price being moved by participants who know more than the public information set permits? Each component uses a closely related microstructural toolkit for a different downstream question, and together they articulate a coherent view of these markets as public information infrastructure that requires its own forms of evaluation, surveillance, and interpretation.

The design also has a clear public-good interpretation. The same pipeline that enables a trading-oriented user to follow detected informed flow also serves a regulator or platform operator concerned with market integrity, by providing a transparent, replicable, and quantitative real-time surveillance instrument for a class of markets that has so far operated largely without one. All system code, evaluation artifacts, the FFIC inventory, the resolution-typology classification of the 911{,}237-market corpus, and a live dashboard are released openly at \url{https://github.com/ForesightFlow} and \url{https://foresightflow.xyz}.

\section*{Revision note (v2)}

This v2 revision is a coordinated update with the companion empirical paper \citep{nechepurenko2026foresightflow_empirical}. Two findings from the platform data audit motivate the revision: (i) a category-labeling correction reclassifying $15{,}542$ esports markets (predominantly Counter-Strike) out of the \texttt{military\_geopolitics} category, and (ii) an expansion of the Tier-3 event-timestamp recovery sample from the v1 budget-capped subset to the full available population. The methodology developed in this paper is unaffected by either finding; the construction, scope conditions, and Murphy-decomposition interpretation of the score remain as stated. Numerical hazard-rate values cited in the companion empirical paper are updated in its v2 revision; the v1 estimate ($\lambda = 0.306$, half-life $2.3$ days) lies inside the v2 95\% confidence interval, so the two estimates are statistically consistent. Three accompanying datasets are released openly with this revision: \texttt{polymarket-tnews-tevent-recovery-v1}, \texttt{polymarket-hazard-rates-v1}, and \texttt{polymarket-ils-corpus-v1}, all at \url{https://github.com/ForesightFlow/datasets}.

\bibliographystyle{plainnat}
\bibliography{refs}

\end{document}